\documentclass[preprint,12pt]{emulateapj-rtx4}
\usepackage[varg]{txfonts}
\usepackage{graphicx}
\usepackage{booktabs}
\usepackage{afterpage}
\usepackage{changepage}
\usepackage{chngcntr}
\bibpunct{(}{)}{;}{a}{}{,}
\usepackage{color}
\usepackage[caption=false]{subfig}
\usepackage[colorlinks=true,citecolor=blue,urlcolor=red]{hyperref}

\usepackage{longtable}

\newlength{\offsetpage}
{\end{adjustwidth}}

\begin{document}

\title{The K2-ESPRINT Project V: a short-period giant planet orbiting a subgiant star$^{\star}$}

\hyphenation{Kepler}

\author{Vincent~Van~Eylen$^{1,2}$, Simon~Albrecht$^{1}$, Davide Gandolfi$^{3,4}$, Fei Dai$^5$, Joshua N. Winn$^5$, Teriyuki Hirano$^6$, Norio Narita$^{7,8,9}$, Hans Bruntt$^1$, Jorge Prieto-Arranz$^{10,11}$, V\'ictor J. S. B\'ejar$^{10,11}$, Grzegorz Nowak$^{10,11}$, Mikkel N. Lund$^{12,1}$, Enric Palle$^{10,11}$, Ignasi Ribas$^{13}$, Roberto Sanchis-Ojeda$^{14,15}$, Liang Yu$^{5}$, Pamela Arriagada$^{16}$, R.~Paul Butler$^{16}$, Jeffrey~D.~Crane$^{17}$, Rasmus Handberg$^{1}$, Hans Deeg$^{10,11}$, Jens Jessen-Hansen$^{1}$, John A.~Johnson$^{18}$, David Nespral$^{10,11}$, Leslie Rogers$^{19}$, Tsuguru Ryu$^{7,8,9}$, Stephen Shectman$^{17}$, Tushar Shrotriya$^{1}$, Ditte Slumstrup$^{1}$, Yoichi Takeda$^{7}$, Johanna Teske$^{16,17}$, Ian Thompson$^{17}$, Andrew Vanderburg$^{18}$, Robert Wittenmyer$^{20,21,22}$}

\affil{$^1$ Stellar Astrophysics Centre, Department of Physics and Astronomy, Aarhus University, Ny Munkegade 120, DK-8000 Aarhus C, Denmark}
\affil{$^2$ Leiden Observatory, Leiden University, 2333CA Leiden, The Netherlands}
\affil{$^3$ Dipartimento di Fisica, Universit\'a di Torino, via P. Giuria 1, I-10125, Torino, Italy}
\affil{$^4$ Landessternwarte K\"onigstuhl, Zentrum f\"ur Astronomie der Universit\"at Heidelberg, K\"onigstuhl 12, D-69117 Heidelberg, Germany}
\affil{$^5$ Department of Physics, and Kavli Institute for Astrophysics and Space Research, Massachusetts Institute of Technology, Cambridge, MA 02139, USA\label{inst2}}
\affil{$^6$ Department of Earth and Planetary Sciences, Tokyo Institute of Technology, 2-12-1 Ookayama, Meguro-ku, Tokyo 152-8551, Japan}
 \affil{$^{7}$ National Astronomical Observatory of Japan, 2-21-1 Osawa, Mitaka, Tokyo 181-8588, Japan}
 \affil{$^{8}$ SOKENDAI (The Graduate University for Advanced Studies), 2-21-1 Osawa, Mitaka, Tokyo 181-8588, Japan}
 \affil{$^{9}$ Astrobiology Center, National Institutes of Natural Sciences, 2-21-1 Osawa, Mitaka, Tokyo 181-8588, Japan}
 \affil{$^10$ Instituto de Astrof\'\i sica de Canarias (IAC), 38205 La Laguna, Tenerife, Spain}
 \affil{$^{11}$ Departamento de Astrof\'\i sica, Universidad de La Laguna (ULL), 38206 La Laguna, Tenerife, Spain}
\affil{$^{12}$ School of Physics and Astronomy, University of Birmingham, Edgbaston, Birmingham, B15 2TT, UK}
\affil{$^{13}$ Institut de Ci\`encies de l'Espai (CSIC-IEEC), Carrer de Can Magrans, Campus UAB, 08193 Bellaterra, Spain}
\affil{$^{14}$ Department of Astronomy, University of California, Berkeley, CA 94720}
\affil{$^{15}$ NASA Sagan Fellow}
\affil{$^{16}$ Department of Terrestrial Magnetism, Carnegie Institution of Washington, 5241 Broad Branch Road, NW, Washington, DC 20015, USA}
\affil{$^{17}$ The Observatories of the Carnegie Institution of Washington, 813 Santa Barbara Street, Pasadena, CA 91101, USA}
\affil{$^{18}$ Harvard-Smithsonian Center for Astrophysics, Cambridge, MA 02138, USA}
\affil{$^{19}$ Department of Astronomy and Division of Geological and Planetary Sciences, California Institute of Technology, MC249-17, 1200 East California Boulevard, Pasadena, CA 91125, USA}
\affil{$^{20}$ School of Physics, University of New South Wales, Sydney 2052, Australia}
\affil{$^{21}$ Australian Centre for Astrobiology, University of New South Wales, Sydney 2052, Australia}
\affil{$^{22}$ Computational Engineering and Science Research Centre, University of Southern Queensland, Toowoomba, Queensland 4350, Australia}

\altaffiltext{$\star$}{Based on observations made with the NOT telescope under programme ID. 50-022/51-503, 50-213(CAT), 52-201 (CAT), 52-108 (OPTICON), 51-211 (CAT),
and ESOs $3.6$\,m telescope at the La~Silla~Paranal~Observatory under programme ID 095.C-0718(A).}

\email{vaneylen@strw.leidenuniv.nl}

\shorttitle{A short-period giant planet orbiting a subgiant star}
\shortauthors{Van Eylen et al.}

\received{receipt date}
\revised{revision date}
\accepted{acceptance date}

\begin{abstract}
{We report on the discovery and characterization of the transiting planet K2-39b (EPIC~206247743b). With an orbital period of 4.6 days, it is the shortest-period planet orbiting a subgiant star known to date. Such planets are rare, with only a handful of known cases. The reason for this is poorly understood, but may reflect differences in planet occurrence around the relatively high-mass stars that have been surveyed, or may be the result of tidal destruction of such planets.
K2-39 is an evolved star with a spectroscopically derived stellar radius and mass of $3.88^{+0.48}_{-0.42}~\mathrm{R_\odot}$ and $1.53^{+0.13}_{-0.12}~\mathrm{M_\odot}$, respectively, and a very close-in transiting planet, with $a/R_\star = 3.4$. Radial velocity (RV) follow-up using the HARPS, FIES and PFS instruments leads to a planetary mass of $50.3^{+9.7}_{-9.4}~\mathrm{M_\oplus}$. In combination with a radius measurement of $8.3 \pm 1.1~\mathrm{R_\oplus}$, this results in a mean planetary density of $0.50^{+0.29}_{-0.17}$ g~cm$^{-3}$. We furthermore discover a long-term RV trend, which may be caused by a long-period planet or stellar companion. 
Because K2-39b has a short orbital period, its existence makes it seem unlikely that tidal destruction is wholly responsible for the differences in planet populations around subgiant and main-sequence stars.
Future monitoring of the transits of this system may enable the detection of period decay and constrain the tidal dissipation rates of subgiant stars.
}
\end{abstract}


\maketitle

\section{Introduction}
In comparison to main-sequence stars, subgiant and giant stars have a higher observed occurrence of exoplanets but have fewer close-in giant planets \citep{bowler2010,johnson2010,reffert2015}. To explain the lack of close-in planets orbiting these stars, there are currently two main theories. In one scenario, close-in planets are destroyed by tidal evolution: they spiral into their host stars as they transfer angular momentum, a process that is expected to be stronger for evolved stars than for main-sequence stars \citep[e.g.][]{rasio1996,villaver2009,schlaufman2013}. 
In another scenario, the lower occurrence rate of short-period gas giant planets orbiting evolved stars is a result of the systematically higher mass of the observed evolved stars compared to the observed main-sequence stars. The shorter lifetime of the inner protoplanetary disks around these more massive stars causes the lower occurrence rate of gas giant planets at short orbital periods \citep[e.g.][]{burkert2007,kretke2009,currie2009}.

Detections of planets around evolved stars are challenging because of additional noise sources in the stellar Radial Velocity (RV) signal \citep[see e.g.][]{reffert2015}, and because the larger stellar radii result in shallower planetary transits. There are currently only four evolved stars ($R \geq 3.5~\mathrm{R_\odot}$) known to host short-period ($\leq 100$ days) transiting planets.

One example is Kepler-91b \citep{lillobox2014}, whose validity as a genuine planet was debated \citep[e.g.][]{sliski2014} until RV confirmation ruled out false positive scenarios \citep{lillobox2014rv,barclay2015}. Kepler-56 is host to two short-period transiting planets \citep{huber2013kepler56}. Kepler-391 is likely an evolved star with two short-period planets (7 and 20 days) that were statistically validated \citep{rowe2014}. Finally, Kepler-432b is an eccentric Jupiter-sized planet orbiting its giant star in 52 days \citep{ciceri2015,quinn2015,ortiz2015}.

Here, we report on the discovery and characterization of K2-39b (EPIC~206247743b), a transiting planet in a 4.6 day orbit around a subgiant star, making it the shortest period planet orbiting such a star known to date. Its transits were observed by the K2 mission \citep{howell2014} in Campaign~3. The transits in this system have also been recently reported by \cite{vanderburg2015firstyear}, who assigned it the status of `planetary candidate'. We conducted radial velocity follow-up observations using HARPS \citep{mayor2003}, FIES \citep{telting2014}, and PFS \citep{crane2010}, which result in a $5\sigma$ measurement of the mass, both confirming the planetary nature of the system and constraining its bulk density. This work is part of the \textit{Equipo de Seguimiento de Planetas Rocosos INterpretando sus Tr\'ansitos} (ESPRINT) project \citep[see][]{sanchisojeda2015,vaneylen2015k2,hirano2015}.

In Section~\ref{sec:observations}, we describe the observations used in this work. In Section~\ref{sec:modeling} we describe the way these data were modeled. In Section~\ref{sec:results} we present the results and in Section~\ref{sec:conclusion} we discuss and conclude.

\section{Observations}
\label{sec:observations}

\subsection{Photometry}
\label{sec:photometry}

The K2 observations \citep{howell2014} are extracted from the raw pixel files, detrended, reduced, and searched for planets following the procedure outlined in \cite{vaneylen2015k2} and using the pipeline publicly available on GitHub\footnote{\url{https://github.com/vincentvaneylen}}. We summarize the important features here. 

\begin{figure}[!htbp]
\centering
\resizebox{\hsize}{!}{\includegraphics{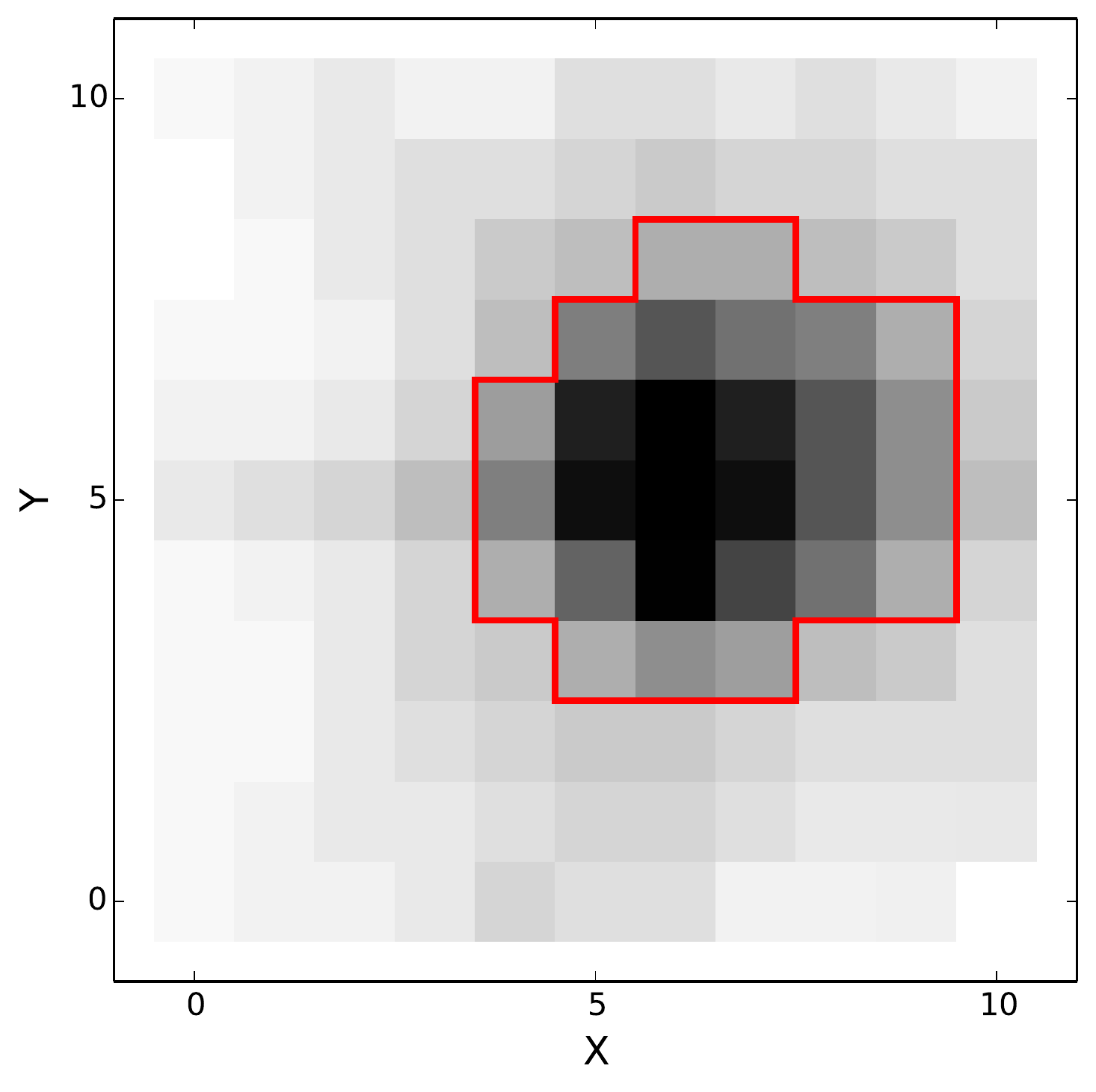}}
\caption{Pixel mask for K2-39. The grey scale indicates the electron count, going from black (high) to white (low). The red line encircles the aperture used to generate the photometry, which includes all pixels with counts more than six times the median flux value.\label{fig:aperture}}
\end{figure}

The aperture that was used to generate a light curve for K2-39 is shown in Figure~\ref{fig:aperture}, and includes all pixels that have a flux level that is at least six times the median flux value in the pixel mask. The light curves are detrended using a polynomial fit of time $T$ and flux $F$ to the centroid positions ($X_c$ and $Y_c$), to remove instrumental effects. Specifically we fit the model $M$, with fitting parameters $t_i$, $x_i$, $y_i$, and $z_1$:
 
 \vspace{-1em}
\begin{eqnarray*}
   M = t_0 + t_1 T + x_1 X_c + x_2 X_c^2 
   + y_1 Y_c + y_2 Y_c^2  + z_1 X_c Y_c.
\end{eqnarray*}

\begin{figure}%[!htbp]
\centering
\resizebox{\hsize}{!}{\includegraphics{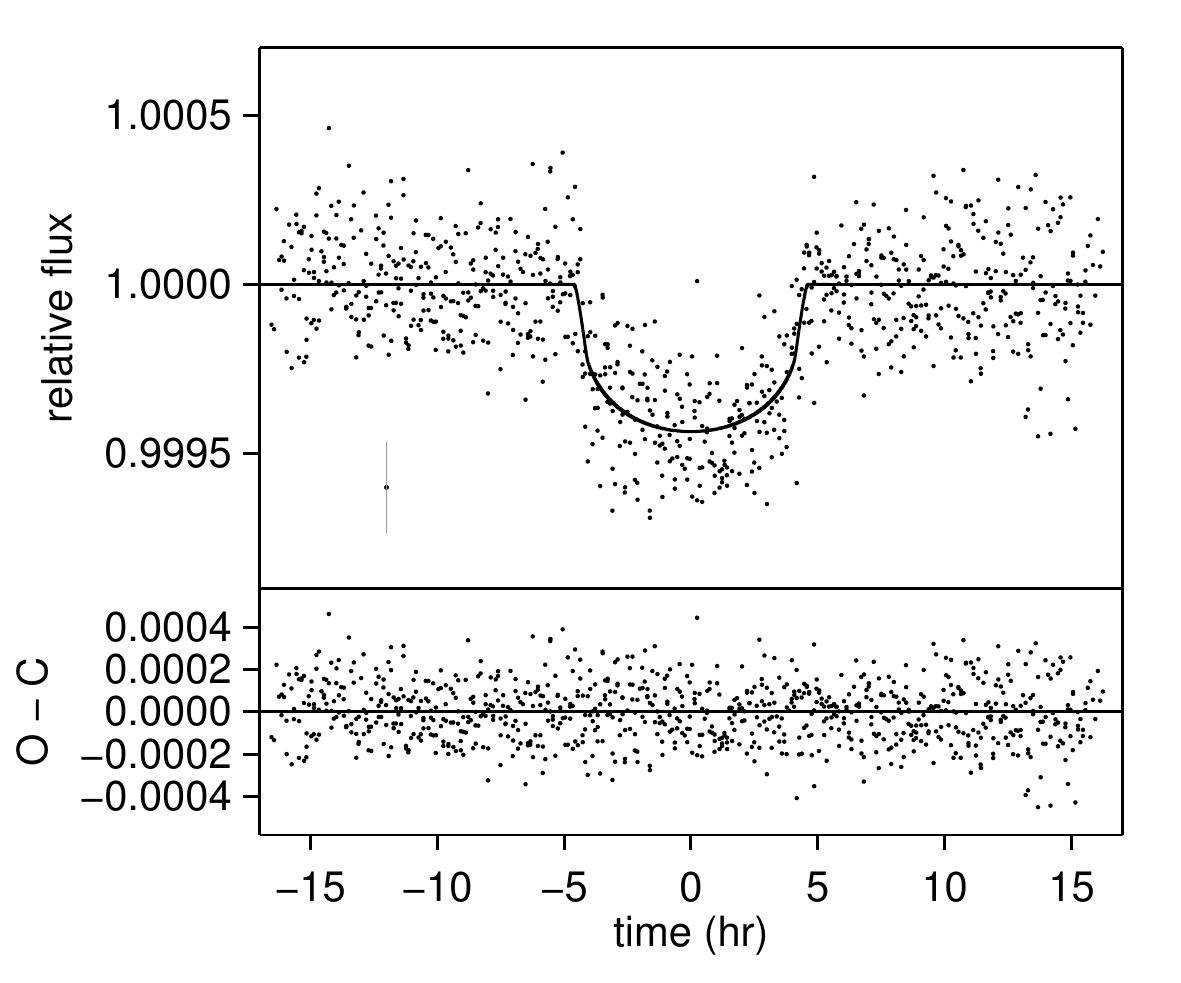}}
\caption{Reduced and phase-folded K2 transit photometry. The best fitted model is shown with a solid line, as well as the residuals after subtracting the model. Error bars are omitted for clarity, but a representative error bar is shown in the top panel.\label{fig:transit}}
\end{figure}

We note that this is a lower-order polynomial than was used by \cite{vaneylen2015k2}, which we found to result in higher-quality photometry in this case. The light curve was fitted in chunks of 650 data points each. We also compared the resulting light curve with one obtained following \cite{sanchisojeda2015}, and found it to be consistent. An initial orbital period is determined using a box least-square (BLS) search algorithm \citep[e.g.][]{kovacs2002}, which is later refined during the fitting procedure (see Section~\ref{sec:orbital}). The final, phase-folded transit light curve is shown in Figure~\ref{fig:transit}.

%%%%%%%%%%%%%%%%%%%%%%%%%%%%%%%%%%%%%%%%%%%%%%%%%%%%%%%%%%%%%%%
%%%%%%%%%%%%%%%%%%%%%%%%%%%%%%%%%%%%%%%%%%%%%%%%%%%%%%%%%%%%%%%

%%%%%%%%%%%%%%%%%%%%%%%%%%%%%%%%%%%%%%%%%%%%%%%%%%%%%%%%%%%%%%%
%%%%%%%%%%%%%%%%%%%%%%%%%%%%%%%%%%%%%%%%%%%%%%%%%%%%%%%%%%%%%%%

\subsection{Imaging follow-up observations}
\label{sec:imaging}

The photometric aperture of K2-39 contains many pixels ({28 in total,} see Figure~\ref{fig:aperture}) and each K2 pixel spans $3.98 \times 3.98$ arcsec. This implies that ground-based imaging is needed to assess the presence of nearby, contaminant stars that may be associated with the system or aligned by chance. 

\begin{figure}%[!htbp]
\centering
\resizebox{\hsize}{!}{\includegraphics{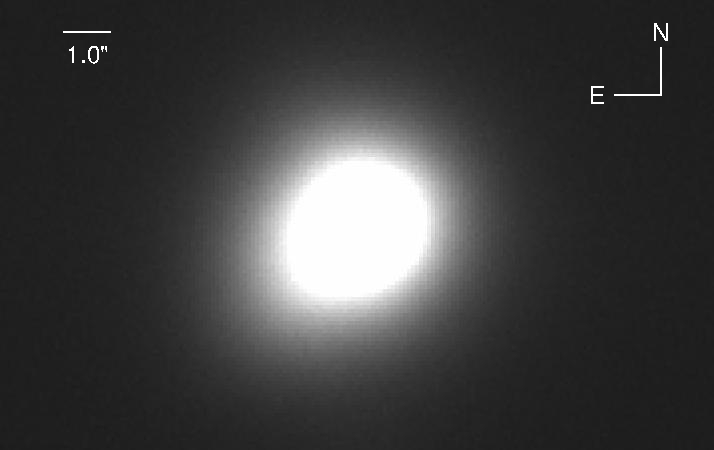}}\\
\resizebox{0.495\hsize}{!}{\includegraphics{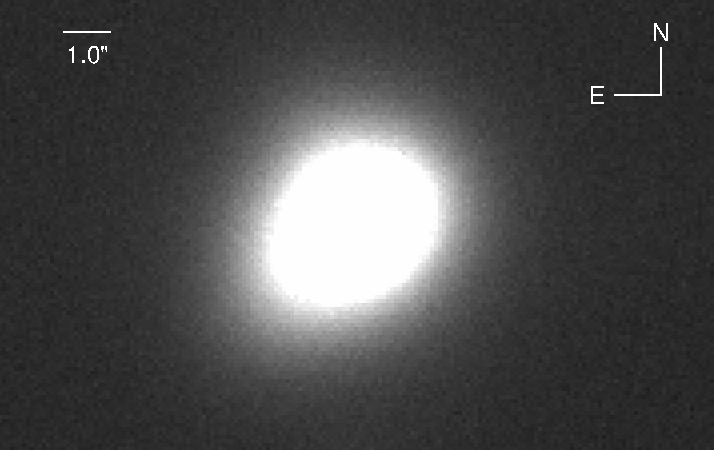}}
\resizebox{0.495\hsize}{!}{\includegraphics{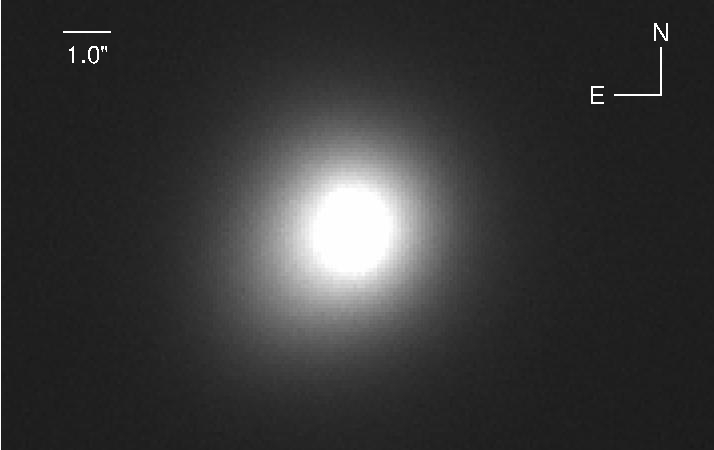}}
\caption{Top: shift-and-add FastCam image of K2-39, constructed by combining the 50\% best individual TCS/FastCam frames. The orientation is North-up and East-right. Bottom: stars observed just before (left) and just after (right) the target, during the same night. The image on the left shows the same distortion, indicating the distortion is likely caused by instrument defocus.\label{fig:fastcam}}
\end{figure}

We performed lucky imaging observations with the FastCam camera \citep{oscoz2008} at the 1.55-meter Telescopio Carlos S\'anchez (TCS). FastCam is a very low noise and fast readout speed EMCCD camera with 512 $\times$ 512 pixels (with a physical pixel size of 16 microns, a scale of 42.5 mas per pixel, and a FoV of $21.2'' \times 21.2''$) and it is cooled down to -90$^\circ$C. On July 30th, 10,000 individual frames of K2-39 were collected in the I-band, with an exposure time of 50~ms for each frame. In order to construct a high resolution, long-exposure image, the individual frames were bias-subtracted, aligned and co-added. In Figure~\ref{fig:fastcam}, we present a high resolution image that was constructed by co-adding the best 50\% of images, so that it has a 5~sec total exposure time. 

%%%%%%%%%%%%%%%%%%%%%%%%%%%%%%%%%%%%%%%%%%%%%%%%%%%%%%%%%%%%%%%
%%%%%%%%%%%%%%%%%%%%%%%%%%%%%%%%%%%%%%%%%%%%%%%%%%%%%%%%%%%%%%%

We find no evidence for a contaminant star within the field of view. The target star shows a deviation from spherical symmetry. To assess if it may be instrumental in nature, we looked at other targets observed during the same night. The target observed just before K2-39 shows the same elongated shape (see Figure~\ref{fig:fastcam}), indicating that the cause of the asymmetry is likely instrumental in nature, due to a defocus.

\begin{figure}%[!htbp]
\centering
\resizebox{\hsize}{!}{\includegraphics{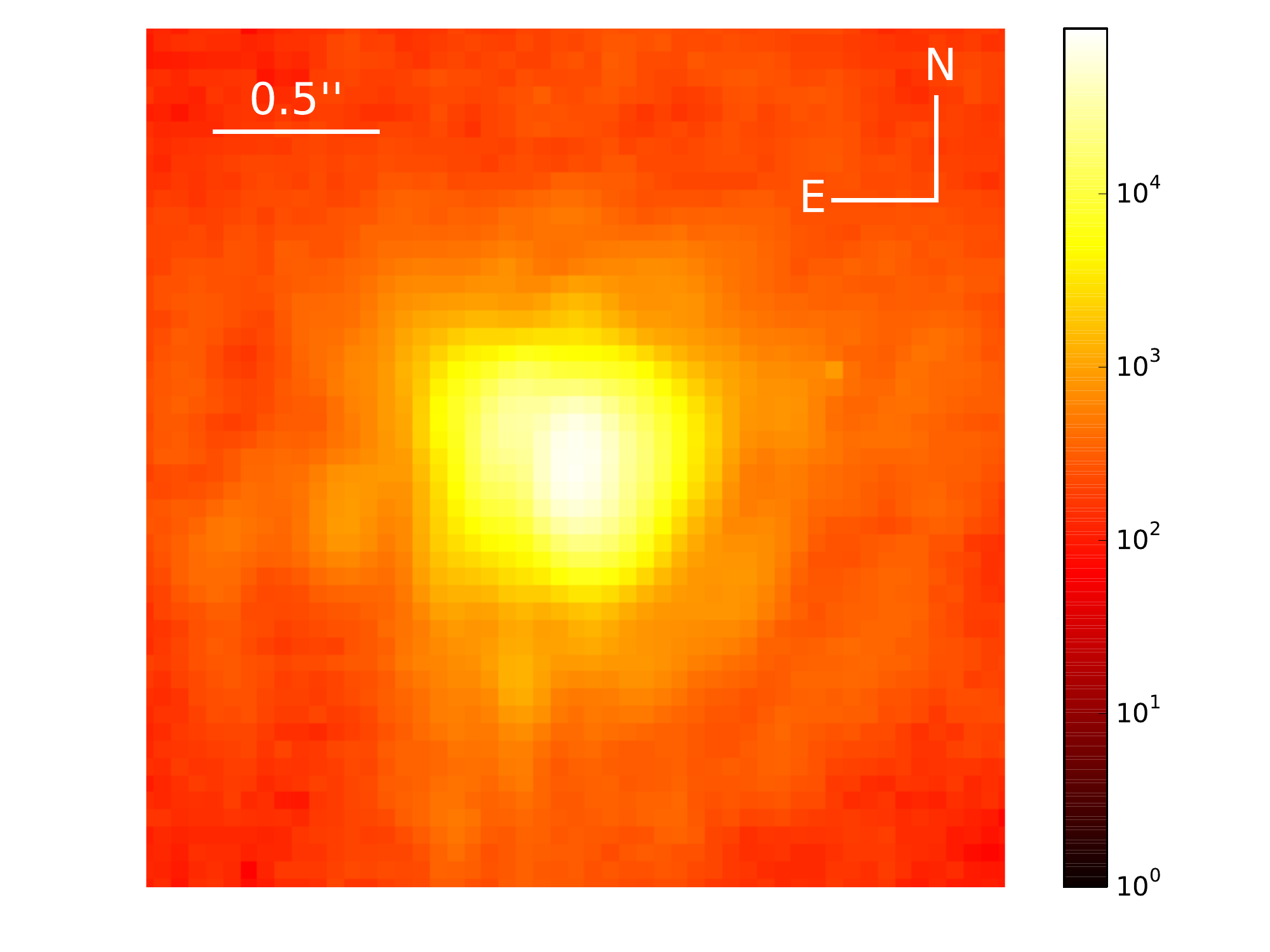}}\\
\resizebox{\hsize}{!}{\includegraphics{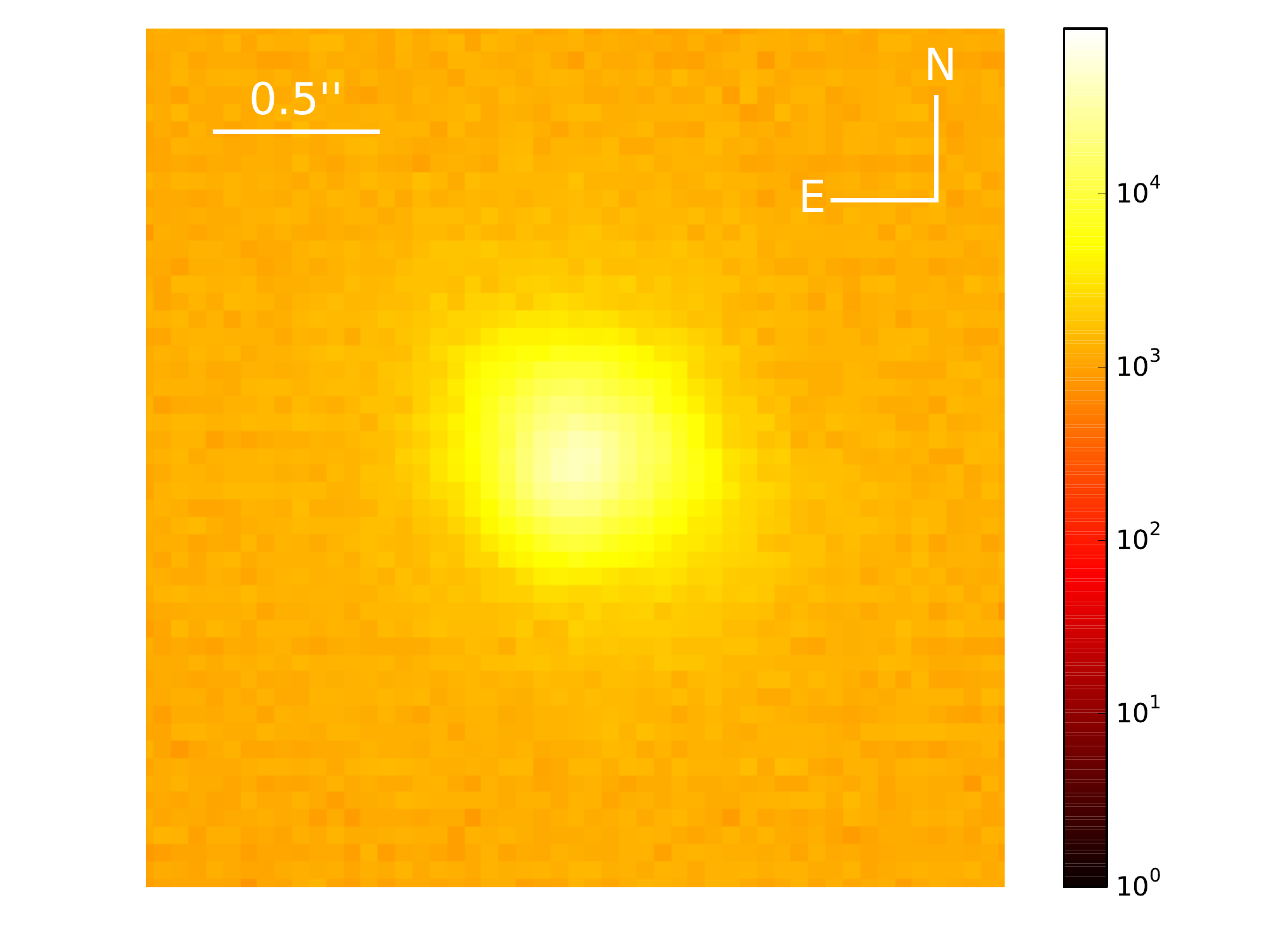}}\\
\resizebox{\hsize}{!}{\includegraphics{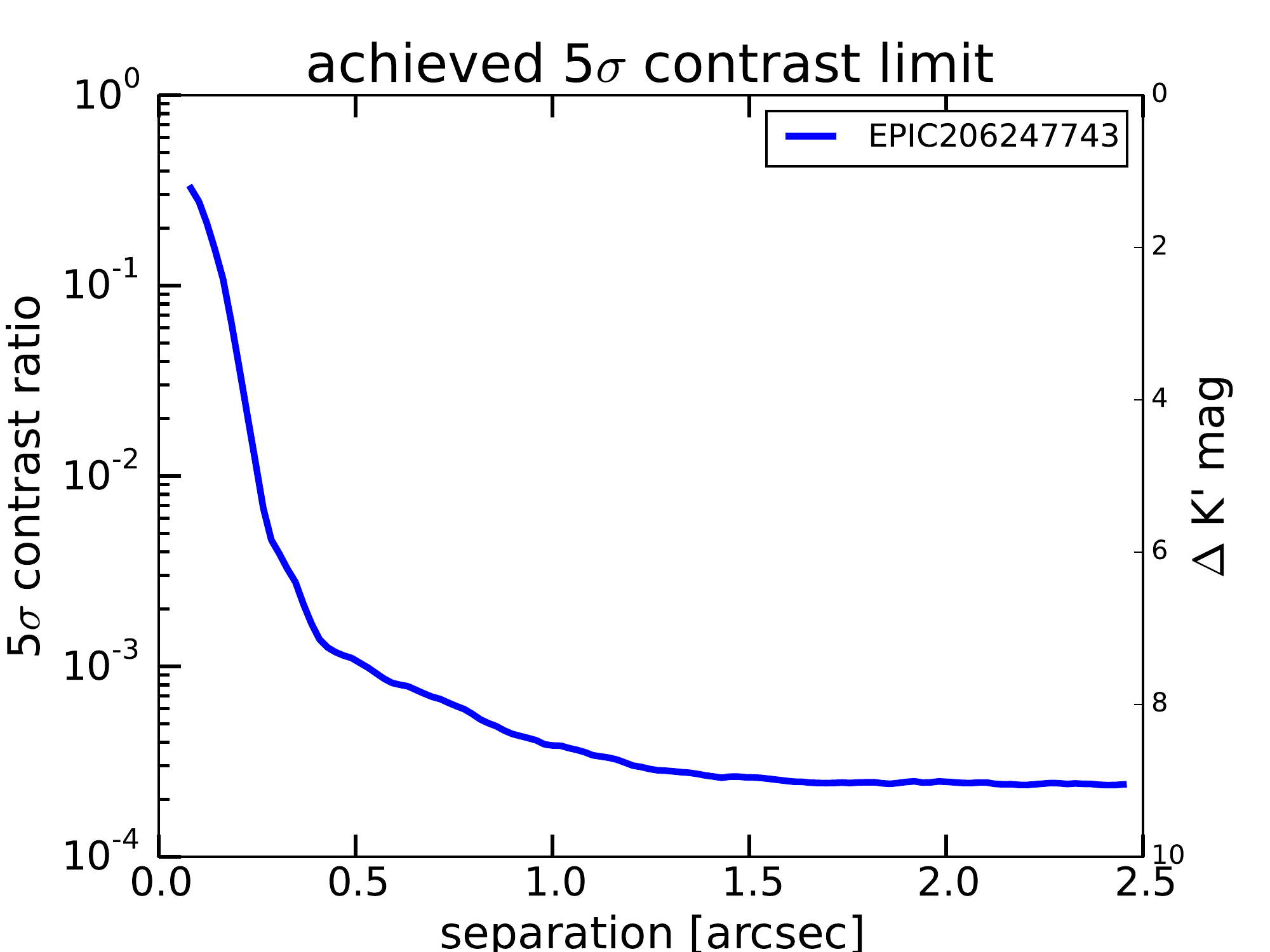}}
\caption{Adaptive optics image using the Subaru telescope. The orientation is North-up and East-left, as in Figure~\ref{fig:fastcam}. In the top figure, the image for K2-39 is shown. The middle figure shows a standard star {(FS151)} observed during the same night, which has a slightly non-circular shape, suggesting a degraded AO performance. The bottom image shows the $5\sigma$ contrast ratio the image provides.\label{fig:aoimage}}
\end{figure}

We further gathered an Adaptive Optics (AO) image using the Subaru telescope{'s Infrared Camera and Spectrograph (IRCS)}, which is shown in Figure~\ref{fig:aoimage} together with the achieved $5\sigma$ contrast limits. The seeing without AO was estimated at 0.4 arcsec, but a cirrus clouds may have degraded the AO performance at the 0.1 arcsec level, as suggested by the PSF of a standard star (FS151) observed during the same night {(see Figure~\ref{fig:aoimage})}. 

%%%%%%%%%%%%%%%%%%%%%%%%%%%%%%%%%%%%%%%%%%%%%%%%%%%%%%%%%%%%%%%
%%%%%%%%%%%%%%%%%%%%%%%%%%%%%%%%%%%%%%%%%%%%%%%%%%%%%%%%%%%%%%%

The AO image reveals no companion objects{, but shows a distortion to the North-East of about $0.2$ arcsec.} The standard star FS151 shows a distortion as well, suggesting the cause for this effect may be instrumental. This asymmetry in the AO observations does not follow the same orientation as the asymmetry in the lucky imaging, and does not have the same scale, so that they are unlikely to have the same origin.

An alternative explanation for the observed asymmetry in the AO image is the presence of a very nearby companion star. If this is the case, such a companion would influence our measurement of the planetary radius, by diluting the transit light curve. {Assuming the distortion is caused by a stellar companion, we can estimate its flux by fitting two two-dimensional Gaussian functions to the image. This method is not ideal because the `companion star' is not well-separated from the target star, but after trying different fitting methods and Gaussian function parameters, we consistently find it to be 4-7 times fainter than the target star. Assuming the companion is associated with the target star, so that it has the same distance, this implies that the companion star would likely be less evolved than the target star, to explain the lower flux contribution. As discussed in detail in Section~\ref{sec:activity}, we find no evidence of such a star (or any other star) in the spectroscopic observations. 
Therefore, we proceed here under the assumption that no companion star is present. We caution that, if there is indeed a nearby star, this would influence the derived planetary parameters.}

\subsection{Spectroscopic follow-up observations}
\label{sec:spectroscopy}

We carried out high-precision RV follow-up observations of K2-39 using the HARPS, FIES, and PFS instruments. 

We started our observations using the FIES spectrograph \citep{frandsen1999,telting2014} mounted at the 2.56-m Nordic Optical Telescope (NOT) of Roque de los Muchachos Observatory (La Palma, Spain). We used the FIES \texttt{high-res} mode (R\,$\approx$\,67\,000) and collected 17 high-resolution spectra from July 2015 until January 2016. We set the exposure time to 15-20 minutes, which resulted in an average signal-to-noise ratio (SNR) of 40 per pixel at 5500\,\AA. We acquired long-exposed ($T_\mathrm{exp}$ $\sim$30\,sec) ThAr spectra right before and after each science exposure to trace the RV drift of the instrument. We used the method by \cite{gandolfi2015} to analyze the data.

{Between 21 August and 13 September 2015,} we observed the system using the HARPS spectrograph at the ESO 3.6-m telescope at La Silla. We acquired 7 high-resolution spectra, using an exposure time of 10 minutes per data point. At order 50, they have an average SNR of 30. The HARPS observations were analyzed using the standard data reduction pipeline \citep{baranne1996,pepe2000}.

We also acquired data using the PFS at the Magellan II Telescope at Las Campanas Observatory. Between and 23 August and 4 September 2015 we obtained 6 high-resolution spectra. Each exposure lasted 20 minutes and resulted in a SNR of 80-100.
PFS uses the iodine technique for calibration and radial velocities were derived using an updated version of the algorithm outlined in \cite{butler1996}.

All RV data points and their observation times are listed in Table~\ref{tab:rvdata}.

\section{Modeling}
\label{sec:modeling}

\subsection{Stellar parameters}
\label{sec:stellar}

We co-added the HARPS spectra and determined the atmospheric parameters following \cite{takeda2002}. We find that the effective temperature $T_\mathrm{eff} = 4881 \pm 20$ K, surface gravity $\log g = 3.44 \pm 0.07$ (cgs), metallicity [Fe/H] = 0.32 $\pm$ 0.04, and a microturbulent velocity of $\xi=0.97 \pm 0.11$ km s$^{-1}$, based on the measurement of equivalent widths of iron lines and on the excitation and ionization equilibria. Following \citet{hirano2012}, we also derive the stellar rotation velocity by fitting the combined HARPS spectrum to obtain $v\sin i_\star = 2.0 \pm 0.5$ km s$^{-1}$. 

\begin{figure}[htb]
\centering
\resizebox{\hsize}{!}{\includegraphics{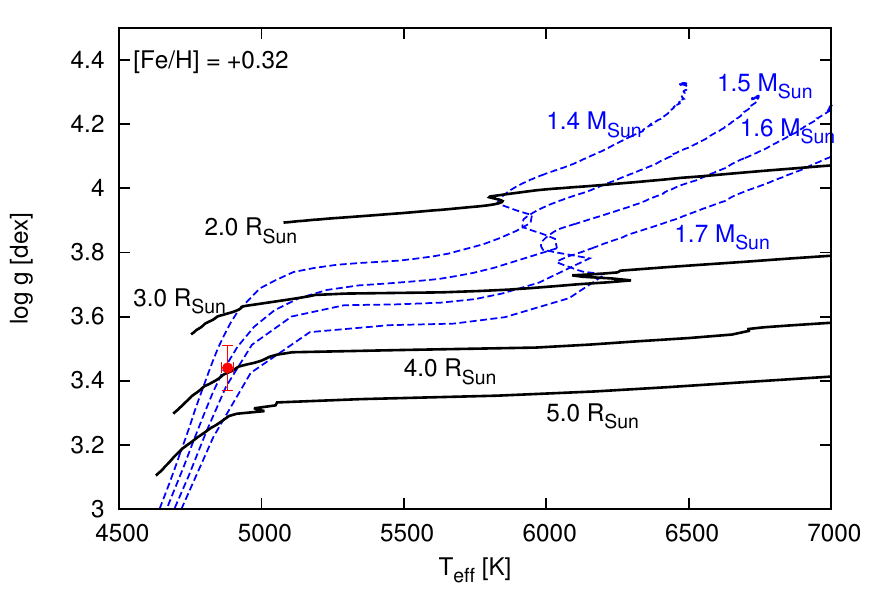}}
\caption{{{$Y^2$ isochrones for [Fe/H] = 0.32, where the blue lines represent the evolution tracks for different stellar masses, and the black curves are curves of constant radius. K2-39 is shown as the red point with its error bars.}}
\label{fig:isochrones}}
\end{figure}

From the stellar atmospheric parameters, we then determine the stellar physical parameters using $Y^2$ isochrones \citep{yi2001}, as shown in Figure~\ref{fig:isochrones}. K2-39 is found to be a metal-rich subgiant star ($M_\star = 1.53^{+0.13}_{-0.12}~\mathrm{M_\odot}$, $R_\star = 3.88^{+0.48}_{-0.42}~\mathrm{R_\odot}$) with an age of $3.09^{+0.92}_{-0.70}$ Gyr. 

Since the stellar parameters of evolved stars are known to be sensitive to the adopted isochrones (evolutional tracks), we also checked the consistency of the derived parameters in two ways. Firstly, we also derived stellar atmospheric parameters using the VWA software\footnote{\url{https://sites.google.com/site/vikingpowersoftware/home}} \citep{bruntt2012}. Again using the effective stellar temperature ($T_{\rm eff}$), surface gravity ($\log g$), and metallicity ([Fe/H]) as input, we infered the stellar mass, radius, and age using BaSTI evolution tracks\footnote{\url{http://albione.oa-teramo.inaf.it/}} following the SHOTGUN method \citep{stello2009}. We found the results to be consistent. Secondly, we check the results by employing the empirical relations of \citet{torres2010}. Consequently, we found $M_\star = 1.39^{+0.11}_{-0.10}~\mathrm{M_\oplus}$ and $R_\star = 3.69^{+0.43}_{-0.38}~\mathrm{R_\oplus}$, which agree with the isochrone-based values within $1\sigma$. We adopt the values derived using the 
$Y^2$ isochrones for the remainder of this work, and list these parameters in Table~\ref{tab:parameters}. 

%%%%%%%%%%%%%%%%%%%%%%%%%%%%%%%%%%%%%%%%%%%%%%%%%%%%%%%%%%%%%%%
%%%%%%%%%%%%%%%%%%%%%%%%%%%%%%%%%%%%%%%%%%%%%%%%%%%%%%%%%%%%%%%

\subsection{Asteroseismology}
Because the star is evolved and relatively bright, we searched the frequency power spectrum of the corrected light curve for indications of stellar oscillations. We performed a search for indications of a seismic power excess using frequency-power spectra prepared using corrected light curves from the K2-Photometry-Pipeline \citep[K2P$^2$][]{lund2015} and the KASOC filter \citep{handberg2014}. No such signal was found. This agrees with the spectroscopic parameters, from which we estimate a frequency of maximum power ($\nu_{\rm max}$) of ${\sim}338 \pm 55\, \rm\mu Hz$ using the $\nu_{\rm max}\propto g/\sqrt{T_{\rm eff}}$ scaling relation \citep{brown1991,kjeldsen1995}. This is 
above the Nyquist frequency of $\nu_{\rm Nyq}\approx 283~\mu \rm Hz$ for K2 long-cadence observations, and conforms with the detection limits presented in \cite{stello2014} for K2 observations. 

With an independent estimate of the effective temperature one may use such a non-detection of seismic signal to place a lower-limit on $\log g$ \citep[see][]{campante2014}. {However, without observations in short-cadence, we are limited by the Nyquist frequency,} and can only set a lower limit of $\log g>3.36$ dex. However, the fact that no signal is seen from back-reflected seismic power in the ``super-Nyquist'' regime (i.e., above $\nu_{\rm Nyq}$) could indicate that $\nu_{\rm max}$ is as high as $400~\mu \rm Hz$ \citep{chaplin2014}, hence $\log g\geq3.5$ dex (consistent with findings from spectroscopy, see Section~\ref{sec:stellar}).

\subsection{Orbital and planetary parameters}
\label{sec:orbital}

We derive the planetary parameters following the procedure outlined in detail in \cite{vaneylen2015k2}. We highlight the key points here. 

\paragraph{Photometric model}

We model the planetary transits assuming a constant orbital period {(linear ephemeris)}, without transit timing variations, and using the analytical model by \cite{mandel2002}. The model was binned to 30 minutes to match the finite integration time of the observations (20 hours of observations around each transit were used), and contains the following parameters: orbital period ($P$), mid-transit time ($T_{\rm mid}$), stellar radius divided by semi-major axis ($R_\star  /a$), planetary radius divided by stellar radius ($R_\mathrm{p}/R_\star$), the cosine of the orbital inclination ($\cos i_\mathrm{o}$), and two limb darkening parameter ($u_1$ and $u_2$) which determine a quadratic law.

\paragraph{RV model}

We model the RV observations by fixing the eccentricity to zero and modeling the projected stellar reflex motion ($K_\star$). In addition, we fit for a systemic velocity offset between the different spectrographs ($\gamma_{\rm spec}$). We furthermore include a quadratic drift (using two parameters, $\phi_1$ and $\phi_2$) as a function of time, relative to an arbitary zero point ($t_0$). We tested whether allowing non-zero orbital eccentricity would affect the derived parameters, and found that not to be the case (see also Section~\ref{sec:results_parameters}).
As a result, the RV model we fit for is

\begin{equation}
 \textrm{RV}(t) = \phi_1(t-t_0) + \phi_2(t-t_0)^2 + \gamma_{\rm spec} + \mathrm{RV_{planet}}
\end{equation}

To account for our incomplete knowledge of stellar activity, we add stellar ``jitter'' to the internal uncertainties of the RV points, so that the minimum reduced $\chi^2$ for the data obtained by each spectrograph is close to unity. Note that two data points were observed in-transit, and we assume the star is aligned with the planet in our model. {Even if the star and planet were misaligned, the effect of this would be below the photon noise.}

\paragraph{Prior information}

We place a Gaussian prior with a width of $0.1$ and a center value derived from the tables by \cite{claret2011} on the sum of the limb darkening parameters ($u_1 + u_2$), while holding the difference ($u_1 - u_2$) fixed at the tabulated value. For the K2 bandpass, we find $u_1 = 0.5902$ and $u_2 = 0.1395$, using $T_\mathrm{eff} = 5000$~K, $\log g = 3.5$ (cgs) and [Fe/H] = 0.3. When we try an eccentric fit, the stellar density (see Section~\ref{sec:stellar}) is used as a Gaussian prior and helps constrain $e$ and $\omega$ \citep[see, e.g.,][]{vaneylen2015}. In this case we furthermore assume an eccentricity prior of $\frac{dN}{de} \propto \frac{1}{(1+e)^4} - \frac{e}{2^4}$ \citep[see][]{shen2008}, and require that the orbits of planet and star do not cross, and sample uniformly in $\sqrt{e} \cos\omega$ and $\sqrt{e} \sin\omega$ to avoid a positive bias \citep[see e.g.][]{lucy1971}. All other parameters have flat (uniform) priors.

%%%%%%%%%%%%%%%%%%%%%%%%%%%%%%%%%%%%%%%%%%%%%%%%%%%%%%%%%%%%%%%
\begin{figure*}%[!tbh]
%\centering
\includegraphics[width=0.5\textwidth]{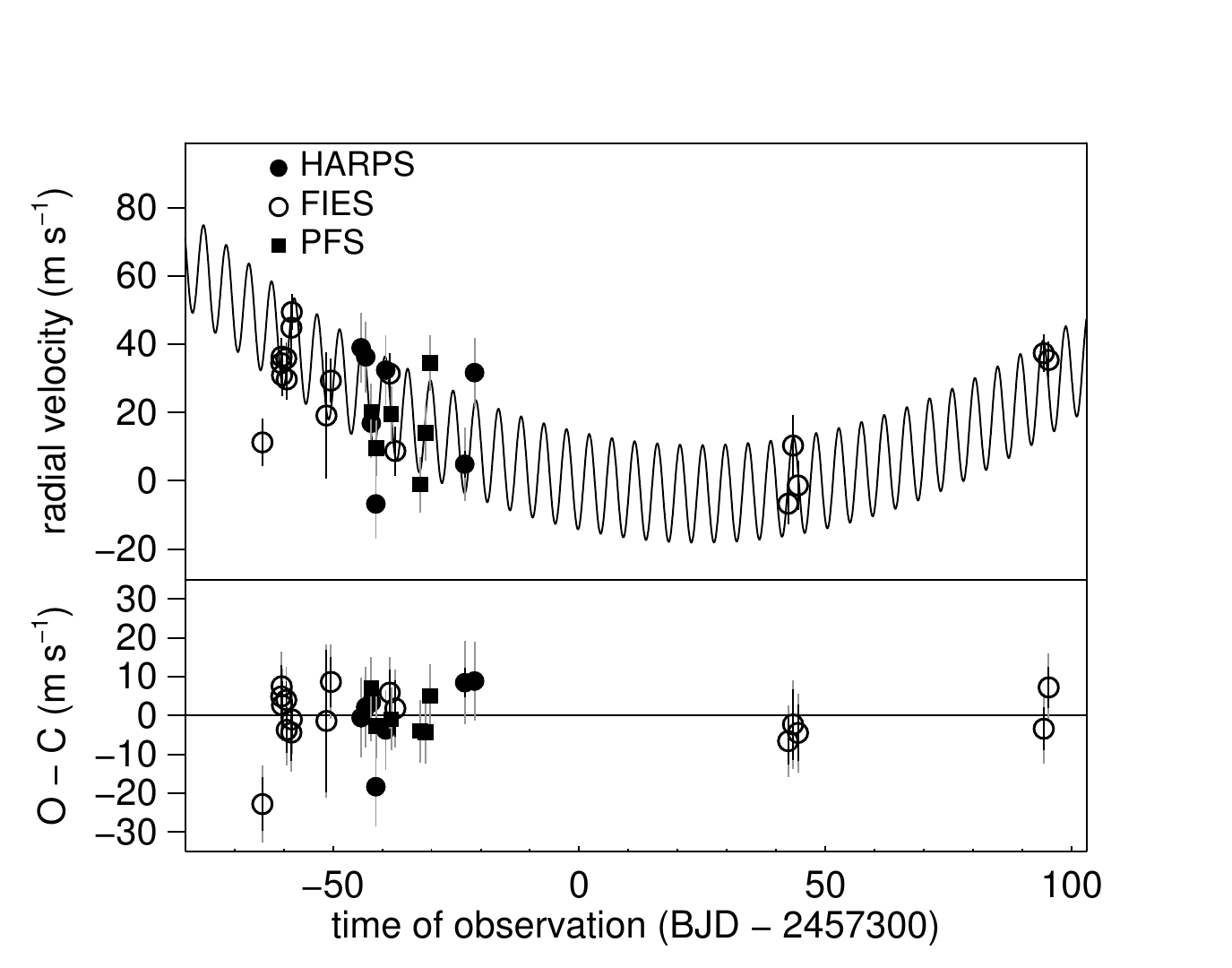}
\includegraphics[width=0.48\textwidth]{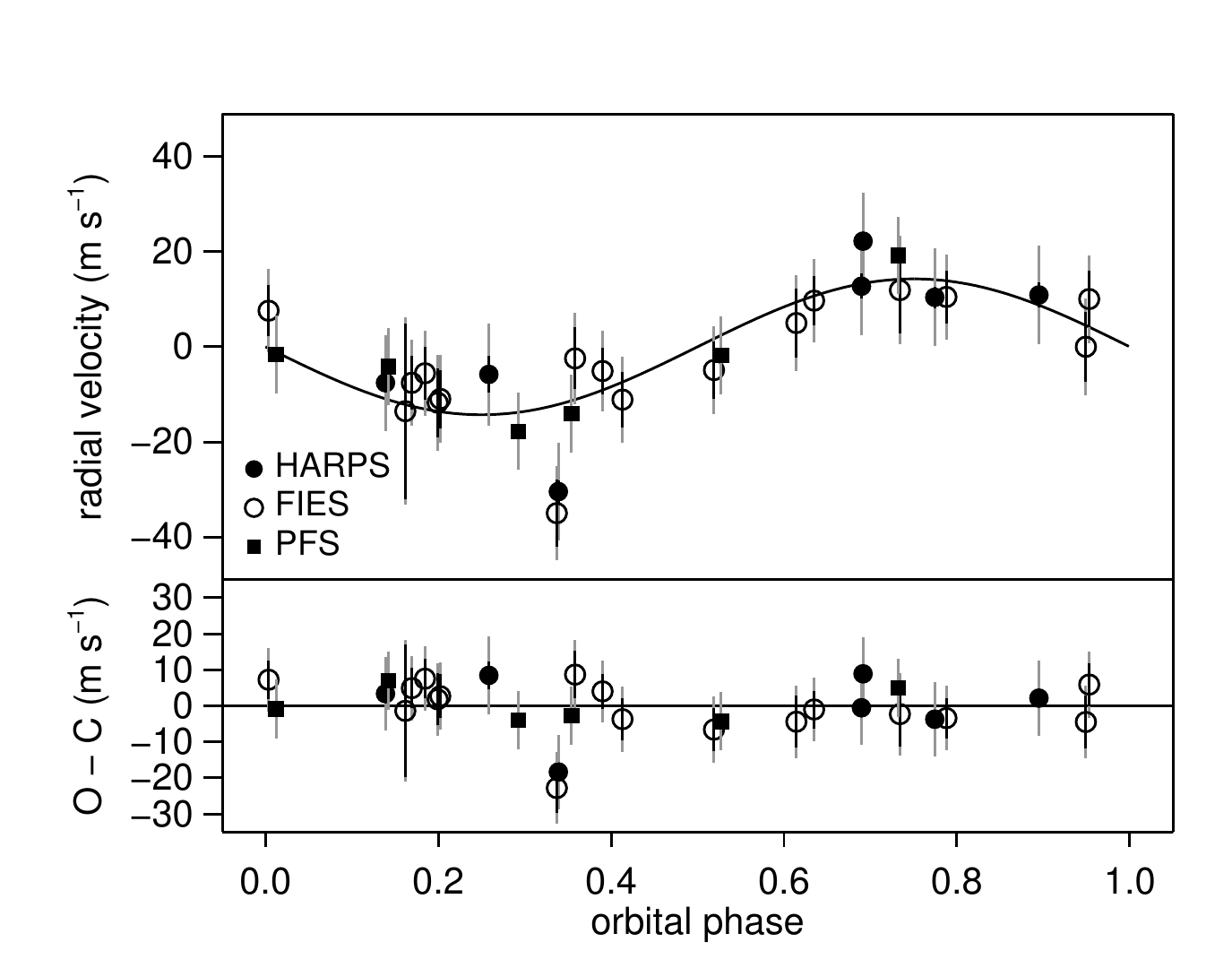}
\caption{RV observations over time (left) and phased (right). The best fitting model is shown with a solid line as well as the residuals after subtracting the model. The internal RV uncertainties are indicated by the black error bars; the gray error bars include an additional ``stellar jitter'' term as explained in the text. \label{fig:rvs}}
\end{figure*}
%%%%%%%%%%%%%%%%%%%%%%%%%%%%%%%%%%%%%%%%%%%%%%%%%%%%%%%%%%%%%%%

\paragraph{Parameter estimation}

We model the planetary transit and the stellar RV signal simultaneously using a Markov Chain Monte Carlo (MCMC) method \citep{tegmark2004}, and following the Metropolis-Hastings algorithm. We run three chains of $10^6$ steps each, with a step size adjusted to obtain an acceptance rate of approximately 25\%. We employed a burn-in phase of $10^4$ points, which were removed from each chain prior to the analysis. We checked for convergence using the Gelman and Rubin diagnostic \citep{gelman1992}. 

The chains are merged and uncertainty intervals encompassing $68.3$\% of the total probability are calculated by excluding the $15.85$\% quantiles on both sides, while median values are used as best estimates. All parameters are reported in Table~\ref{tab:parameters}.

\begin{table*}[htb]
  \begin{center}
    \caption{System parameters \label{tab:parameters}}
    \smallskip
    \begin{tabular}{l  r@{$\pm$}l }
      \tableline\tableline
      \noalign{\smallskip}
      Parameter &  \multicolumn{2}{c}{K2-39 (EPIC 206247743)}  \\
      \noalign{\smallskip}
      \hline
      \noalign{\smallskip}
      \multicolumn{3}{c}{Basic properties} \\
      \noalign{\smallskip}
      \hline
      \noalign{\smallskip}
      2MASS ID	 				& \multicolumn{2}{c}{22332842-0901219} \\
      Right Ascension				& \multicolumn{2}{c}{22 33 28.414}  	\\ 
      Declination				& \multicolumn{2}{c}{-09 01 21.97}  	\\
      Magnitude (\textit{Kepler})		& \multicolumn{2}{c}{10.58}  		\\
      Magnitude ($V$)				& \multicolumn{2}{c}{10.83}		\\
      \hline
      \noalign{\smallskip}
      \multicolumn{3}{c}{Stellar parameters from spectroscopy} \\
      \noalign{\smallskip}
      \hline
      \noalign{\smallskip}
      Effective Temperature, $T_{\rm_{eff}}$ (K) 			& $4881$ & $20$ 	\\
      Surface gravity, $\log g$ (cgs)            			& $3.44$ & $0.07$ 	\\
      Metallicity, [Fe/H]                        			& $0.32$ & $0.04$ 	\\
      Microturbulence (km\,s$^{-1}$)             			& $0.97$ & $0.11$ 	\\
      Projected rotation speed, $v \sin i_{\star}$ (km\,s$^{-1}$) 	& $2.01$ & $0.50$  \\
      Assumed Macroturbulence, $\zeta$ (km\,s$^{-1}$) 			& $2.61$ & $0.39$  \\
      Stellar Mass,   $M_{\star} $ ($M_{\odot}$)			& \multicolumn{2}{c}{$1.53^{+0.13}_{-0.12}$}\\
      Stellar Radius, $R_{\star} $ ($R_{\odot}$)	 		& \multicolumn{2}{c}{$3.88^{+0.48}_{-0.42}$} \\
      Stellar Density, $\rho_\star$ (g cm$^{-3}$)$^{\rm a}$ 		& $0.036$ & $0.011$ 		 \\
      Age (Gyr)								& \multicolumn{2}{c}{$3.09^{+0.92}_{-0.70}$}	\\
      \hline
      \noalign{\smallskip}
      \multicolumn{3}{c}{Fitting (prior) parameters} \\
      \noalign{\smallskip}
      \hline
      \noalign{\smallskip}
      Limb darkening prior $u_1 + u_2$ 					& $0.73$ & $0.1$ 		\\
      Stellar jitter term HARPS (m\,s$^{-1}$) 				& \multicolumn{2}{c}{10}  	\\
      Stellar jitter term FIES (m\,s$^{-1}$) 			& \multicolumn{2}{c}{7}  	\\
      Stellar jitter term PFS (m\,s$^{-1}$) 				& \multicolumn{2}{c}{8}  	\\
      \hline
      \noalign{\smallskip}
      \multicolumn{3}{c}{Adjusted Parameters from RV and transit fit} \\
      \noalign{\smallskip}
      \hline
      \noalign{\smallskip}
                                %   16              2              6
      Orbital Period, $P$ (days) 					& $4.60543$  & $0.00046$ 			\\
      Time of mid-transit, $T_{\rm mid}$ (BJD$-2450000$) 		& $6980.8236$ & $ 0.0039$  			 \\[3pt]
      Orbital Eccentricity, $e$ 					& \multicolumn{2}{c}{$0$ (fixed)} 	 \\[3pt]
      Cosine orbital inclination, $\cos i_{\rm o}$ 			& \multicolumn{2}{c}{0.167$^{+0.075}_{-0.069}$}	 \\[3pt]
      Scaled Stellar Radius, $R_\star/a$ 				& \multicolumn{2}{c}{$0.293^{+0.045}_{-0.030}$}	 \\[3pt]
      Fractional Planetary Radius, $R_{\rm p}/R_\star$ 			& \multicolumn{2}{c}{$0.01925^{+0.00099}_{-0.00076}$}  \\ [3pt]
      Linear combination limb darkening parameters (prior \& transit fit), $u_1+ u_2$, & $0.773$ & $0.083$ 		 \\ [3pt]
      Stellar Density (prior \& transit fit), $\rho_\star$ (g\,cm$^{-3}$) & \multicolumn{2}{c}{$0.036^{+0.014}_{-0.012}$}	 \\ [3pt]
      Stellar radial velocity amplitude, $K_\star$ (m\,s$^{-1}$)      	& \multicolumn{2}{c}{$14.4^{+2.6}_{-2.6}$}  \\ [3pt]
      Linear RV term, $\phi_{1}$ (m\,s$^{-1}$/day) 			& $-0.313$ & $0.052$  				 \\
      Quadratic RV term, $\phi_{2}$ (m\,s$^{-1}$/day) 			& $0.0063$ & $0.0012$ 				\\
      Systemic velocity HARPS, $\gamma_{\rm HARPS}$ (km\,s$^{-1}$) 	& $24.4688$ & $0.0052$  				 \\
      Systemic velocity FIES,   $\gamma_{\rm FIES}$ (km\,s$^{-1}$)   	& $24.5458$ & $0.0056$ 				 \\
      Systemic velocity PFS,   $\gamma_{\rm PFS}$ (km\,s$^{-1}$)    	& $-0.0196$ & $0.0044$  \\
      \noalign{\smallskip}
	\hline
      \noalign{\smallskip}
      %Assumed Macroturbulence, $\zeta$ (km\,s$^{-1}$) & \multicolumn{6}{c}{$2$}  \\
      \multicolumn{3}{c}{Indirectly Derived Parameters} \\
      \noalign{\smallskip}
      \hline
      \noalign{\smallskip}
      Impact parameter, b                           			& \multicolumn{2}{c}{$0.57^{+0.15}_{-0.20}$}	\\ [3pt]
      Planetary Mass,   $M_{\rm p} $ ($M_{\oplus}$)$^{\rm b}$ 		& \multicolumn{2}{c}{$50.3^{+9.7}_{-9.4}$} 	 \\ [3pt]
      %\textbf{Mass upper limit (95\% confidence)}, $M_{\rm p} $ ($M_{\oplus}$)$^{\rm b}$ & \multicolumn{2}{c}{$12.0$} 	 \\ [3pt]
      Planetary Radius, $R_{\rm p} $ ($R_{\oplus}$)$^{\rm b}$ 		& \multicolumn{2}{c}{$8.2^{+1.1}_{-1.1}$} 	 \\[3pt]
      Planetary Density, $\rho_{\rm p}$ (g\,cm$^{-3}$) 			& \multicolumn{2}{c}{$0.50^{+0.29}_{-0.17}$} \\
      Semi-major axis, $a$ (AU)	 				& \multicolumn{2}{c}{$0.062^{+0.010}_{-0.012}$} \\
      \noalign{\smallskip}
      \tableline
      \noalign{\smallskip}
      \noalign{\smallskip}
      \multicolumn{3}{l}{{\sc Notes} ---}\\
      \multicolumn{3}{l}{$^{\rm a}$ This value is used as a prior during the fitting procedure.}\\
      \multicolumn{3}{l}{$^{\rm b}$ Adopting an Earth radius of $6371$~km and mass of $5.9736\cdot10^{24}$~kg.}\\
      \noalign{\smallskip}
    \end{tabular}
  \end{center}
\end{table*}

\section{Results}
\label{sec:results}

\subsection{Planet confirmation and properties}
\label{sec:results_parameters}

We determine a planetary radius of $8.2^{+1.1}_{-1.1}~\mathrm{R_\oplus}$ and a planet mass of $50.3^{+9.7}_{-9.4}~M_{\oplus}$. We obtain a planetary bulk density of $0.50^{+0.29}_{-0.17}$~g\,cm$^{-3}$. The planet is very close to its star, with $R_\star/a = 0.293^{+0.045}_{-0.030}$. We note that if there is an unseen companion star contaminating the light curve (see Section~\ref{sec:imaging}), the planet would be larger and its density would be lower.

During the fitting procedure, we assumed a circular orbit, because the orbital period of 4.6 days suggests that tidal effects have circularized any initial eccentricity. Out of caution, we also try a solution in which we allow non-circular orbits. We find that a circular orbit is favored, with an upper limit (at 95\% confidence) of $e = 0.24$. The resulting planetary mass, $53.8^{+10}_{-9.9}~\mathrm{M_\oplus}$ is consistent with the circular fit. We adopt the values from the circular solution in Table~\ref{tab:parameters}. The best transit fit is shown in Figure~\ref{fig:transit} and the radial velocity observations are shown in Figure~\ref{fig:rvs}. The RMS values of the RVs from each spectrograph after the best fitting model is subtracted are 8.6~m~s$^{-1}$ (HARPS), 7.3~m~s$^{-1}$ (FIES), and 4.4~m~s$^{-1}$ (PFS).

{For a circular orbit, the transit duration directly constrains the mean stellar density. Following the procedure used by \cite{vaneylen2015} to validate Kepler-449b/c and Kepler-450b/c/d, we find that a transit fit constrains the bulk density of the host star to $[0.026,0.14]$~g~cm$^{-3}$ at 95\% confidence, assuming the planet has a circular orbit. The stellar density derived from the transit for a circular orbit is furthermore fully consistent with the bulk density of K2-39 derived from spectroscopy ($0.036 \pm 0.011$~g~cm$^{-3}$), giving further credibility to the fact that this star is indeed the host of the planet.
More generally, the transit duration provides independent evidence that the planet is orbiting an evolved star. For example, we find that if the planet would orbit a star with a solar mean density ($1.408$~g~cm$^{-3}$), it would require the planet to have an eccentricity in the interval $[0.78,0.94]$ at 95\% confidence. Given the short orbital period, we find such a scenario not to be feasible. } 

We find evidence of a long-period companion, which we modeled as a quadratic trend with $\phi_1 = -0.313 \pm 0.052$ m\,s$^{-1}$\,d$^{-1}$ and $\phi_2 = 0.0063 \pm 0.0012$ m\,s$^{-1}$\,d$^{-1}$, with $t_0 = 2457300$ BJD. We check whether the data warrants the inclusion of both parameters, and find this to be the case. Including the quadratic term, we find $\chi^2 = 552.1$, while only including a linear term $\chi^2 = 579.6$. Calculating the Bayesian Information Criterian (BIC) with 12 and 11 degrees of freedom, respectively, and 30 RV data points, we find that the quadratic term is clearly favored (with a BIC of 593, versus 617 for the linear case). If we count all photometry data points as well, the BIC numbers change but the quadratic term remains clearly favored. Nevertheless, the trend is dependent on the two latest observations, so that more observations are needed to fully interpret it. Assuming the trend is caused by a companion object, it has a period that is longer than the time span of the data. 
As a result, its true orbit and amplitude are difficult to constrain. {We attempted to do so by allowing a second body in the MCMC fit, rather than the quadratic trend, but find that the orbit is consistent with all orbital periods longer than 125 days, and amplitudes corresponding to objects at least as massive as Jupiter. Further observations may help determine if the trend is caused by an additional planet or a self-luminous companion object. We note that the AO image (see Section~\ref{sec:imaging}) suggests there may be a nearby companion star. If we roughly estimate this potential companion to be at an angular distance of about 0.2 arcsec, and use 255~pc as the distance between an observer and the star as estimated by RAVE distance calibrations \citep{kordopatis2013,francis2013}, this would imply a minimum distance of $\sim$50~AU between the two stars. This would imply an orbital period of order hundred years, making it unlikely that the quadratic trend observed in a time span of only a few 
months is caused by such an orbit.}

\subsection{Stellar activity and light blending}
\label{sec:activity}

\begin{figure}[htb]
\centering
\resizebox{\hsize}{!}{\includegraphics{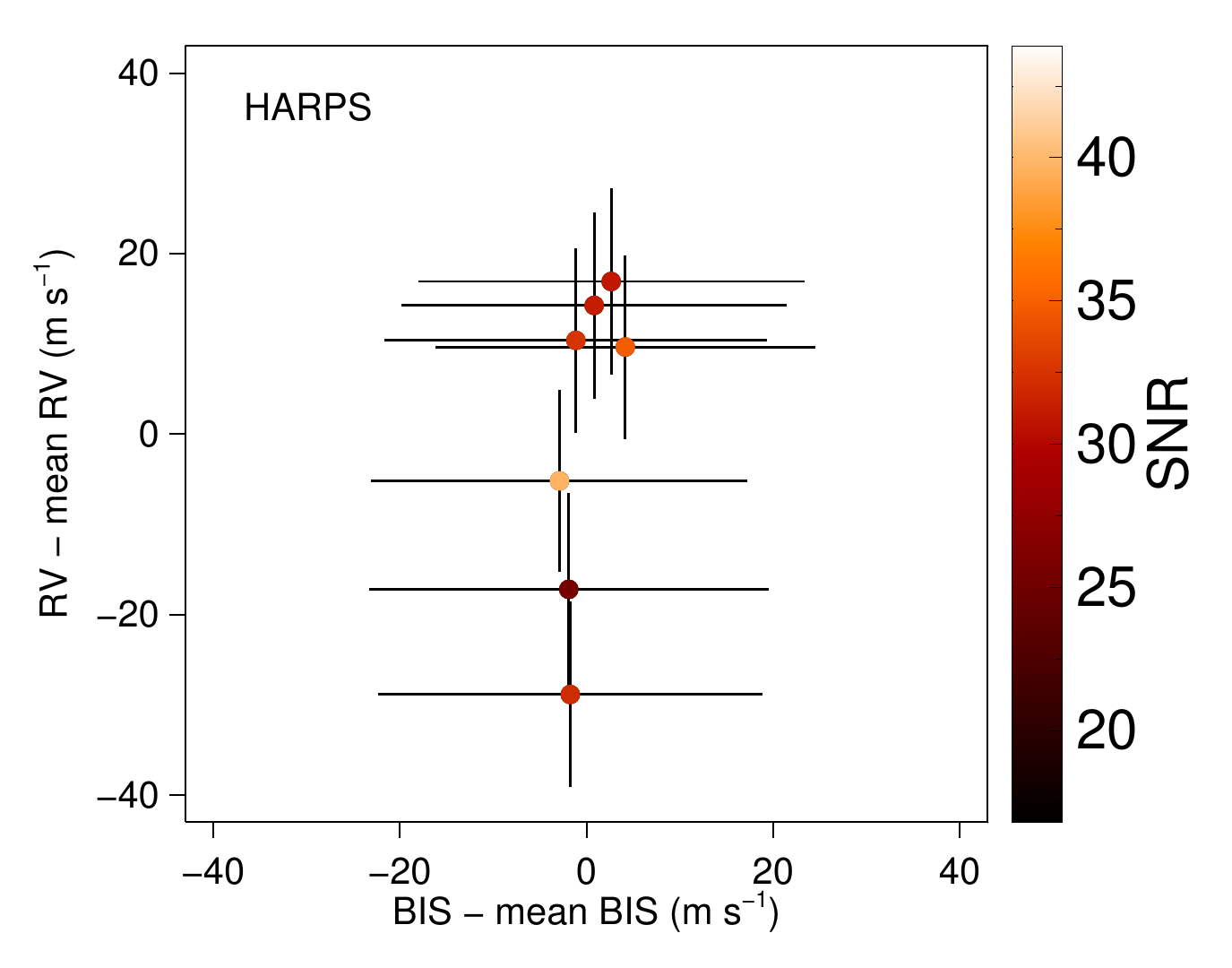}}
\resizebox{\hsize}{!}{\includegraphics{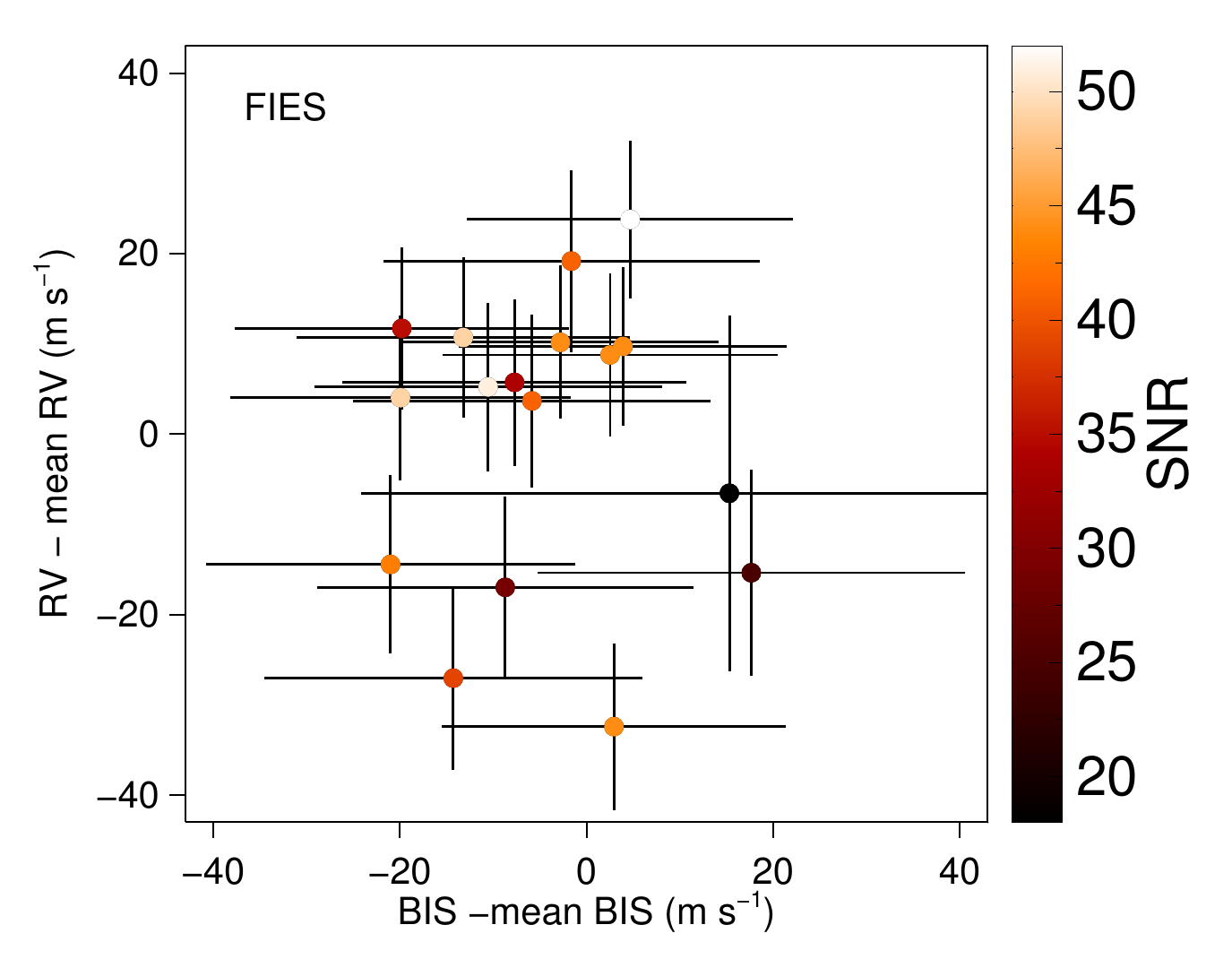}}
\caption{{{Bisectors (BIS) from HARPS (top) and FIES (bottom) CCFS are plotted versus the stellar RVs. The color code indicates the signal-to-noise ratio in the stellar spectra obtained around a wavelength range of $5560$\,\AA. There is no evidence for correlations. The BIS uncertainties are taken to be three times the RV uncertainties. The mean BIS values for HARPS and FIES are 49 m~s$^{-1}$ and -4.6 m~s$^{-1}$, respectively.}}
\label{fig:bisectorplot}}
\end{figure}

We check if any observed RV signal could be caused by stellar activity, by calculating the bisectors (BIS) as defined by \cite{queloz2001} for the HARPS and FIES observations. The results are shown in Figure~\ref{fig:bisectorplot}. We calculate the Pearson correlation coefficient. For HARPS, this is 0.66 with a p-value of 0.11, with 7 data points and 5 degrees of freedom. For FIES, this is -0.03 with a p-value of 0.91, with 17 data points and 15 degrees of freedom. This implies that in neither of the data sets is there any evidence for a correlation at a significance level of 0.01. 

%%%%%%%%%%%%%%%%%%%%%%%%%%%%%%%%%%%%%%%%%%%%%%%%%%%%%%%%%%%%%%%
%%%%%%%%%%%%%%%%%%%%%%%%%%%%%%%%%%%%%%%%%%%%%%%%%%%%%%%%%%%%%%%

%%%%%%%%%%%%%%%%%%%%%%%%%%%%%%%%%%%%%%%%%%%%%%%%%%%%%%%%%%%%%%%
\begin{figure}[htb]
\centering
\resizebox{\hsize}{!}{\includegraphics{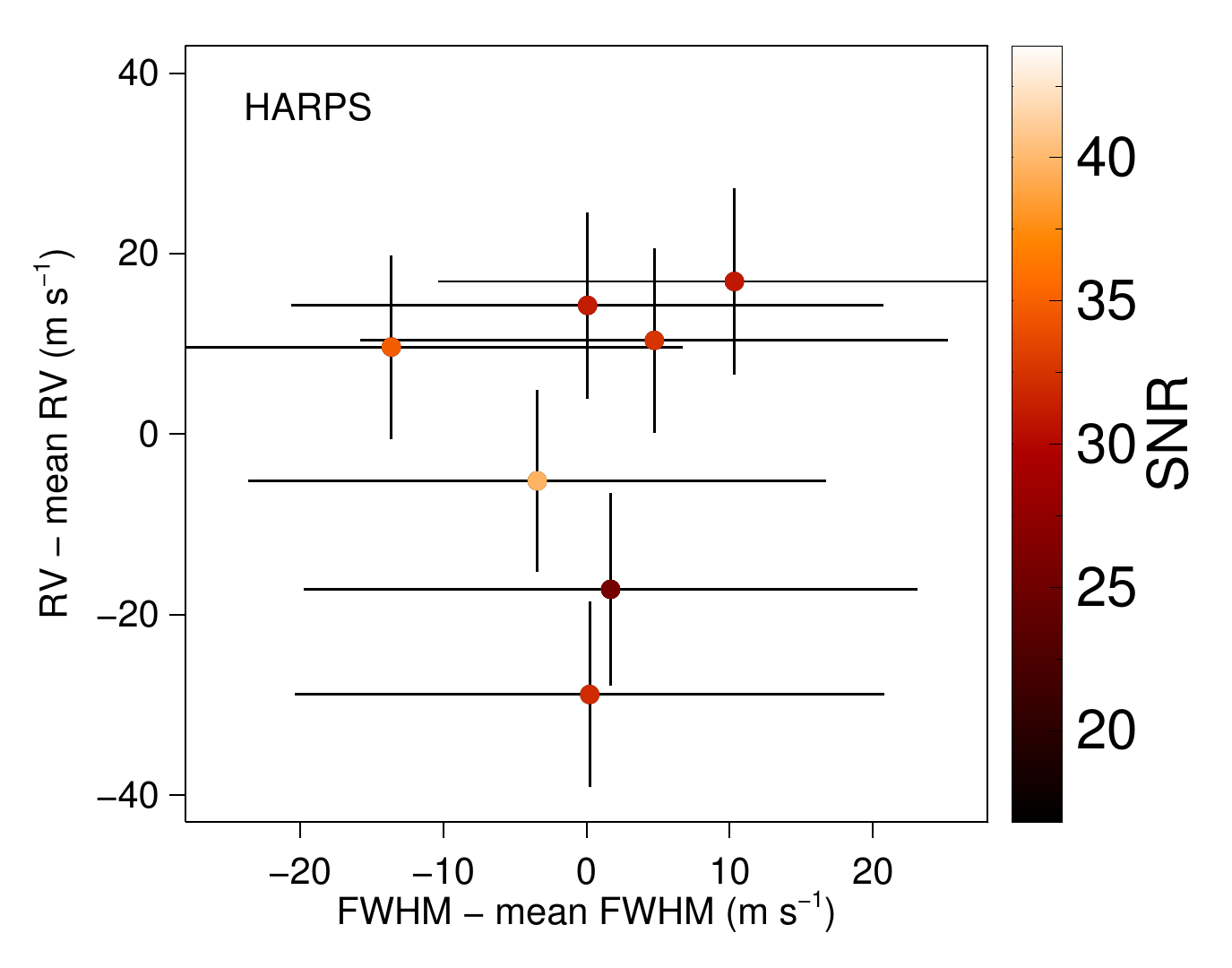}}
\resizebox{\hsize}{!}{\includegraphics{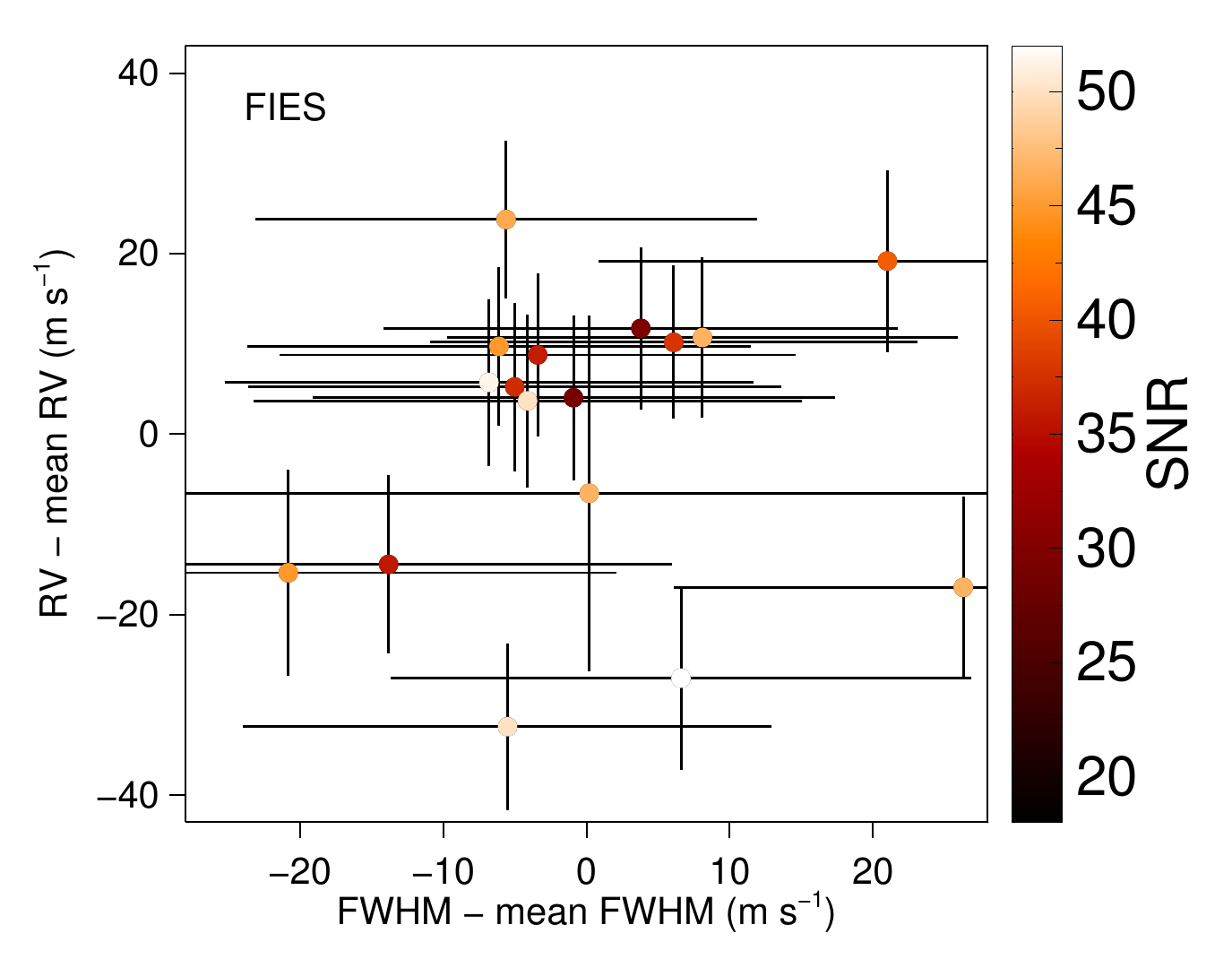}}
\caption{{{Full-width at half maximum (FWHM) from HARPS (top) and FIES (bottom) CCFS are plotted versus the stellar RVs. The color code indicates the signal-to-noise ratio in the stellar spectra obtained around a wavelength range of $5560$\,\AA. There is no evidence for correlations. The mean FWHM values for HARPS and FIES are 6,813 m~s$^{-1}$ and 11,552 m~s$^{-1}$, respectively.}}
\label{fig:fwhm}}
\end{figure}
%%%%%%%%%%%%%%%%%%%%%%%%%%%%%%%%%%%%%%%%%%%%%%%%%%%%%%%%%%%%%%%

We furthermore check if there are any correlations between the Mount Wilson activity index and the RV observations. For the HARPS observations, we have 7 data points and find a Pearson correlation coefficient of 0.23 with a p-value of 0.66, indicating no evidence for correlation. For the PFS data, we have 6 data points and find a Pearson correlation coefficient of 0.24 with a p-value of 0.64. As for the bisector data discussed above, the Mount Wilson measurements show no evidence that the RV variation is caused by stellar activity.

We checked the K2 light curve for evidence of a rotational modulation signal, but could not clearly determine any period of stellar rotation. This may be due to systematic effects in the photometry.

{If an unseen stellar contaminant would have a $v\sin i$ and RV which are very similar to that of K2-39, this may remain undetected in the measured bisectors. However, such a hypothetical companion would still influence the shape of the cross-correlation function (CCF), which can be measured through the Full-Width at Half Maximum \citep[FWHM, see e.g.][]{santerne2015}. We calculate these values for the HARPS and FIES observations and compare them with the RV measurements, as shown in Figure~\ref{fig:fwhm}. For HARPS, we find a Pearson correlation coefficient of 0.10 with a p-value of 0.83, while for FIES we find a Pearson coefficient of 0.12 and a p-value of 0.64, implying there is no evidence for a correlation in either data set. We also checked if the FWHM measurements showed any correlation with the long-term trend seen in Figure~\ref{fig:rvs}, but found no evidence for that either.}

{We further checked the high-resolution spectroscopic observations (see Section~\ref{sec:spectroscopy}) for the presence of a second set of spectral lines, which would be caused by a companion star of a different stellar type. To do so, we looked at the cross-correlation function and found no evidence of any companion star. We furthermore did a visual inspection of the $H_\alpha$ lines for any features caused by a contaminant star, and found no evidence of this. We have also visually inspected the spectrum at 6079-6084 $\AA$, and again found no evidence of any secondary features as deep as $> 10\%$ of the spectrum continuum.}
\subsection{Optical phase curve}

\begin{figure}[!bth]
\centering
\resizebox{\hsize}{!}{\includegraphics{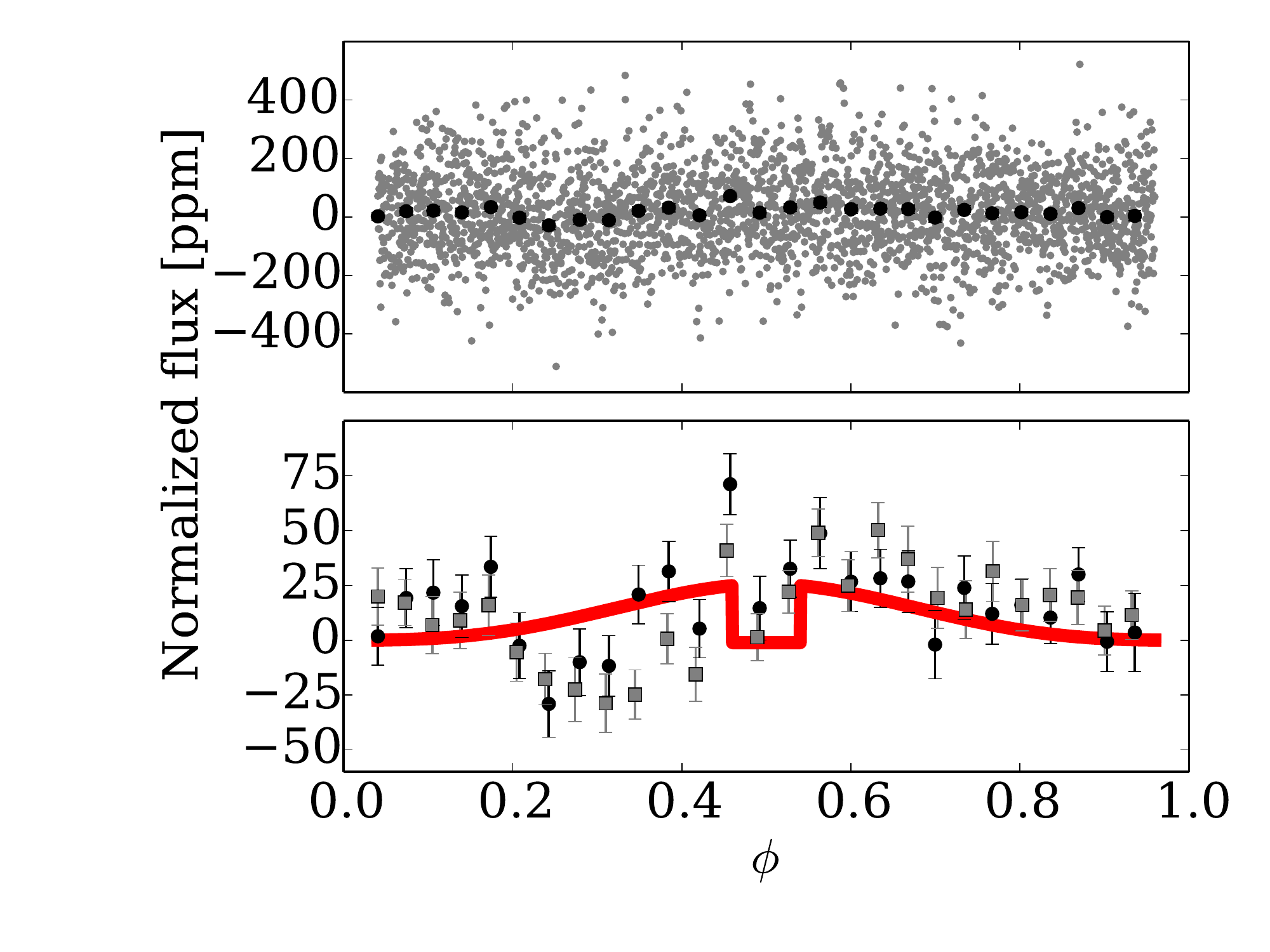}}
\caption{Reduced and phase-folded K2 photometry for K2-39b (transit excluded). \textit{Top:} the K2 observations are shown in grey, with 100-point bins in black circles. \textit{Bottom:} as before, bins in black circles, the best fitted model is shown with a solid red line, and photometry extracted by \cite{vanderburg2015} shown in grey squares. The duration of the occultation is fixed to the duration of the transit, which is the case for circular orbits.\label{fig:phasecurve}}
\end{figure}

The K2 photometry also provides some information on the out-of-transit variation. In general, such variations can be caused by light emitted or reflected by the planet, as well as ellipsoidal modulation of the star caused by the planet, and Doppler beaming \citep[see e.g.][]{esteves2015}. The latter two effects are very small for this system. Given the quality of the data we neglect them here. We model the emitted and reflected light by a Lambert sphere model and fix the nightside temperature to zero. By assuming a circular orbit, we fix the occultation to occur at $\phi = 0.5$ and with a duration equal to that of the primary transit. In this simple model, we fit the light curve (out of the transit and occultation) for a single parameter, the amplitude $F_0$:

\begin{equation}
 F = F_0 \frac{\sin z + (\pi - z) \cos z}{\pi},
\end{equation}

where $z$ is defined as $\cos z = -\sin i \cos(2\pi\phi)$, with $\phi$ describing the orbital phase calculated from mid-transit. {To remove long-term residual trends in the photometry, we run a moving median filter with a width of twice the orbital period before modeling the data.} After doing so, a simple MCMC analysis results in $F_0 = 26.0^{+5.3}_{-5.6}$ ppm and this model is shown in Figure~\ref{fig:phasecurve}. {Such an amplitude would imply a maximum geometric albedo of the planet $A_\mathrm{g}~=~F_0 (a/R_p)^2$ in the interval $[0.64, 0.98]$ within 68\% confidence, or a maximum brightness temperature of the planet of $3050 \pm 100$~K. However, it is clear from the figure that this simple model does not adequately describe the observations, in particular around $\phi = 0.25$. We know of no astrophysical effect that can easily explain the observed dip at this phase, so that the origin is likely instrumental. To check if a different analysis method can avoid this, we compare our data with 
the photometry extracted by \cite{vanderburg2015} using a different method. However, as shown in Figure~\ref{fig:phasecurve}, we find that these data show a similar trend.}

%%%%%%%%%%%%%%%%%%%%%%%%%%%%%%%%%%%%%%%%%%%%%%%%%%%%%%%%%%%%%%%
% %%%%%%%%%%%%%%%%%%%%%%%%%%%%%%%%%%%%%%%%%%%%%%%%%%%%%%%%%%%%%%%
% \begin{figure}%[!tbh]
% \centering
% \resizebox{\hsize}{!}{\includegraphics{phasecurve_interpretation}}
% \caption{Geometric albedo versus planetary (brightness) temperature. The geometric albedo and temperature combinations allowed by the phase curve observations are shown in grey, while the equilibrium temperature is shown in blue.\label{fig:phasecurve_interpretation}}
% \end{figure}
% %%%%%%%%%%%%%%%%%%%%%%%%%%%%%%%%%%%%%%%%%%%%%%%%%%%%%%%%%%%%%%%

%%%%%%%%%%%%%%%%%%%%%%%%%%%%%%%%%%%%%%%%%%%%%%%%%%%%%%%%%%%%%%%
\begin{figure*}%[!tbh]
\centering
\resizebox{0.49\hsize}{!}{\includegraphics{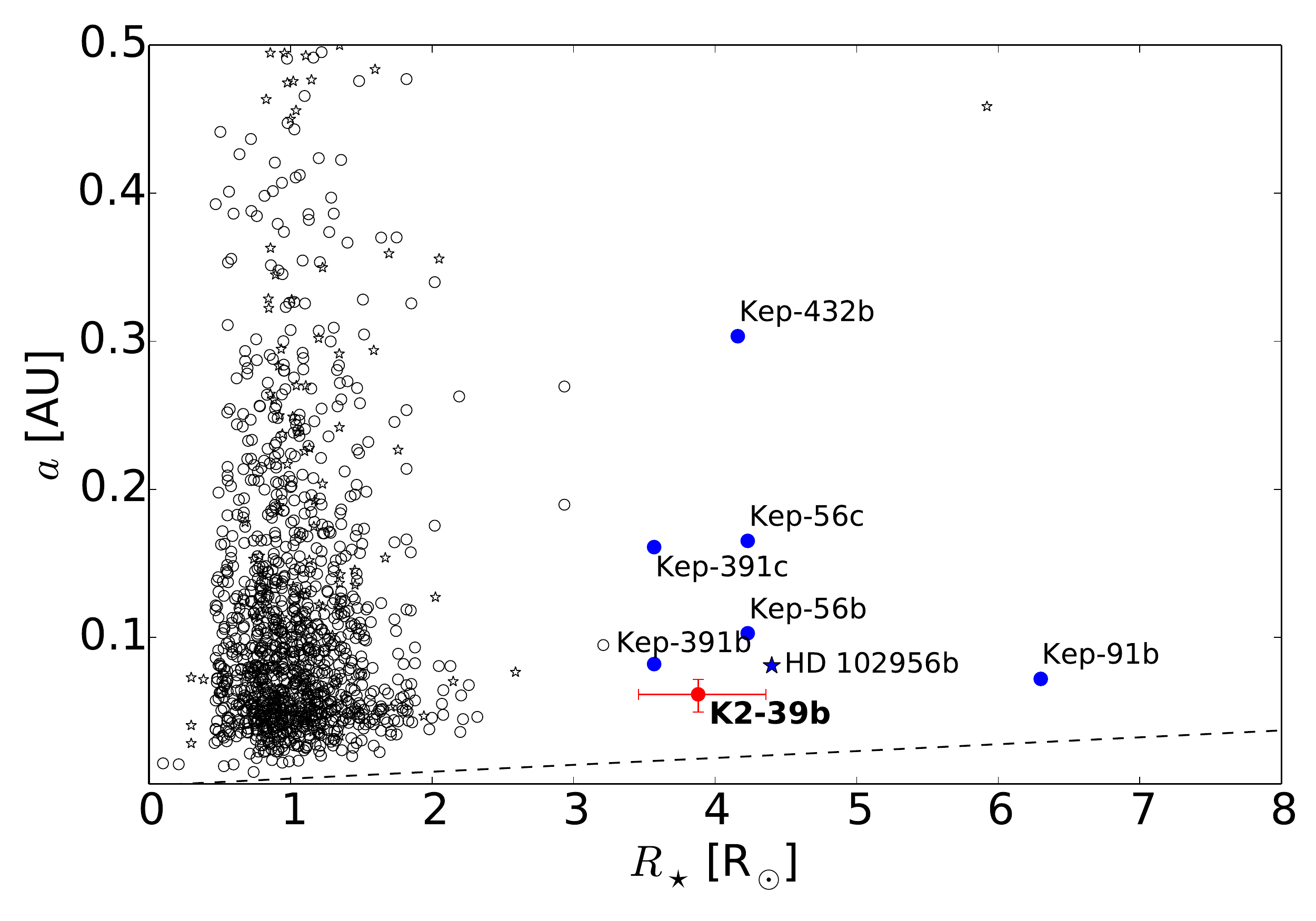}}
\resizebox{0.49\hsize}{!}{\includegraphics{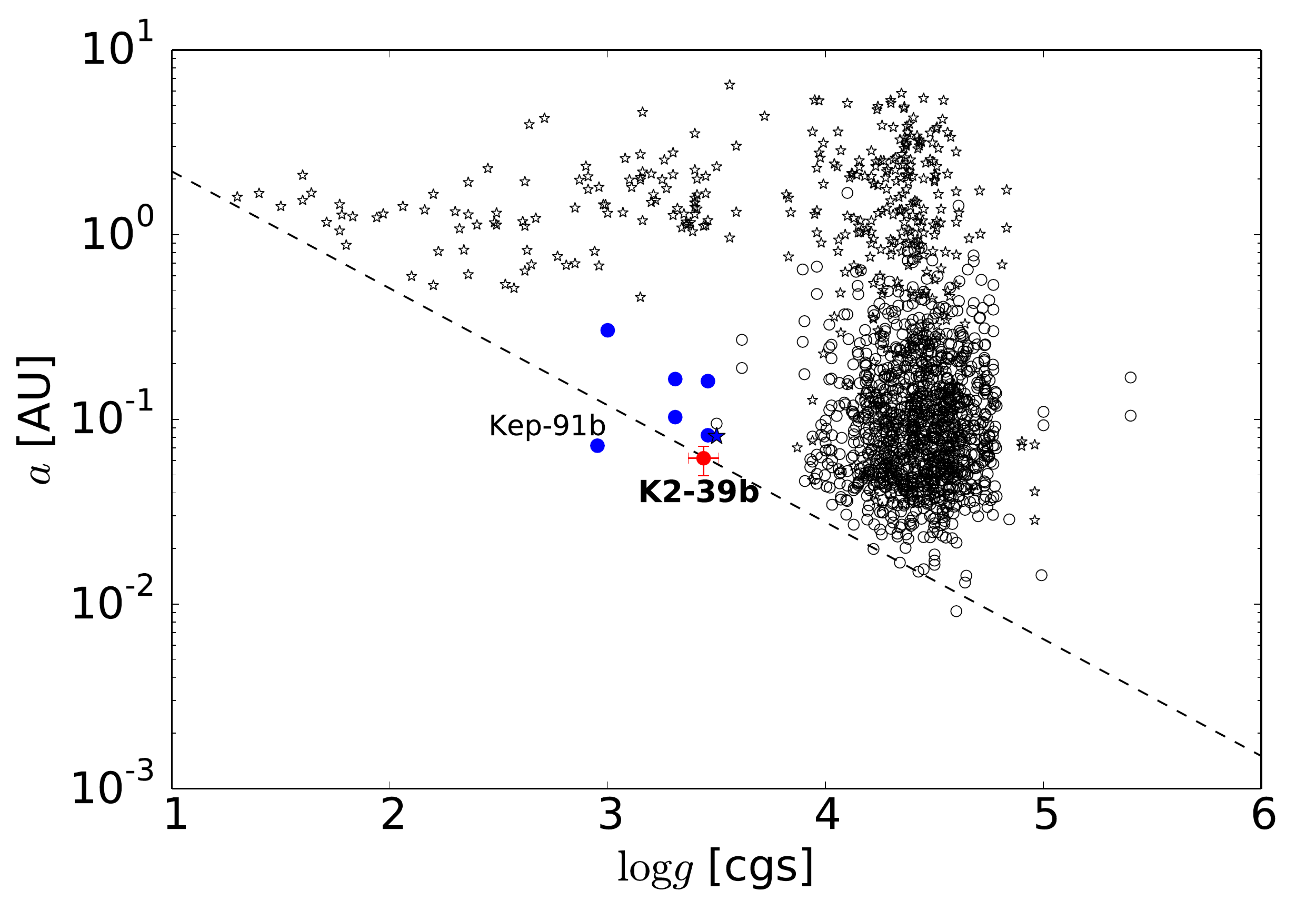}}
\caption{K2-39b versus other confirmed planets, with transiting planets (open circles) and non-transiting planets (open stars) taken from exoplanets.org (accessed on 20 February 2016, error bars omitted for clarity). Short-period transiting planets orbiting evolved stars are indicated with filled blue circles, with values taken from: Kepler-91 \citep{lillobox2014}, Kepler-56 \citep{huber2013kepler56}, Kepler-391 \citep{rowe2014}, Kepler-432 \citep{ciceri2015}, and HD 102956b \citep{johnson2010}. \textit{Left:} stellar radius versus semi-major axis, where the dotted line indicates $R_\star = a$. Of all evolved stars, K2-39 has a planet with the shortest semi-major axis. \textit{Right:} $\log g$ versus semi-major axis. The dotted line indicates the (empirical) border line defined by \cite{nowak2013}. Only Kepler-91b orbits inside of the line, while K2-39b falls exactly on top of it.\label{fig:contextplots}}
\end{figure*}
%%%%%%%%%%%%%%%%%%%%%%%%%%%%%%%%%%%%%%%%%%%%%%%%%%%%%%%%%%%%%%%

%\enlargethispage{2cm}

\section{Discussion and conclusions}
\label{sec:conclusion}

We reported on the discovery and characterization of a giant transiting planet orbiting a subgiant star with a short period. There are only a few known cases of such systems, which are thought to be rare. Of these systems, K2-39b, reported here, has the shortest orbital period. 

By combining the K2 transit photometry with high resolution spectroscopic measurements from HARPS, FIES, and PFS, we are able to measure the planetary mass and mean density. The system shows a long-term quadratic trend indicative of an additional body in the system. The current data do not span enough time to characterize the properties of this body.

{We now discuss the importance of this planet in the context of planet (re)inflation in Section~\ref{sec:reinflation}, and in the context of planet evolution in Section~\ref{sec:discussion_tides}.}

\subsection{(Re)inflation?}
\label{sec:reinflation}

{Measuring the inflation of planets orbiting giant stars is interesting, as it may help distinguish between inflation mechanisms \citep{lopez2016}. \cite{lopez2016} suggest that planets with an orbital period of 10-20 days are likely not inflated while their host star is on the main sequence, but may become inflated as their host star evolves.}

{Despite its low density, K2-39b is not inflated in the sense that its radius is not larger than what would be expected for a pure H/He planet \citep{fortney2007} with this mass. With a mass of $50.3^{+9.7}_{-9.4}$~M$_\oplus$, the planet is likely too small to fall into the regime where inflation is important. With an orbital period of 4.6 days, the planet has received high radiation even when the host star still resided on the main sequence, unless its orbital period used to be longer. We note that the adaptive optics image presented in Figure~\ref{fig:aoimage} cannot rule out a nearby self-luminous companion, although we find no evidence of such a hypothetical companion contributing significant flux in the high-resolution spectroscopic observations. However, if a companion star nevertheless exists, this may affect the planet mass and radius, and its derived mean density. Extreme adaptive optics observations would be needed to rule out such a close companion star.}

\subsection{Tidal evolution}
\label{sec:discussion_tides}

K2-39b joins a small sample of short-period transiting planets orbiting (subgiant) stars. This is illustrated in Figure~\ref{fig:contextplots}, where K2-39b is shown together with Kepler-91b, Kepler-56b/c, Kepler-391b/c, and Kepler-432b. The only non-transiting planet in the same part of the diagram is HD~102956b \citep{johnson2010}. Of all these planets, K2-39b has the lowest semi-major axis and the shortest orbital period. K2-39b is closer to its host star than Kepler-91b, which orbits a more evolved star.  

Because the scarcity of short-period planets orbiting subgiant stars may be a result of tidal destruction \citep{rasio1996,villaver2009,schlaufman2013}, it is interesting to investigate how long K2-39b can survive. Under the assumption that the planet remains in its current orbit, the stellar surface will reach the planet once $R_\star \approx 14~\mathrm{R_\odot}$. Based on the stellar mass of $1.53^{+0.13}_{-0.12}~\mathrm{M_\odot}$, the isochrones suggest this will happen in $150 \pm 90$ Myr. This provides a conservative upper limit on the remaining lifetime of the planet. 

In addition to the evolution of the stellar surface, the planet may spiral inwards as its orbital period decays due to tides. Following \cite{schlaufman2013}, and Equation 11 therein, we can estimate the timescale of orbital decay:

\begin{equation}
 t = 10~\mathrm{Gyr} \frac{Q_\star/k_\star}{10^6} \left(\frac{M_\star}{\mathrm{M_\odot}}\right)^{1/2} \left(\frac{M_\mathrm{p}}{M_\mathrm{Jup}}\right)^{-1} \left(\frac{R_\star}{\mathrm{R_\odot}}\right)^{-5} \left( \frac{a}{0.06\mathrm{AU}}\right)^{13/2}.
\end{equation}

Here, $Q_\star$ is the tidal quality factor of the star, and $K_\star$ its tidal Love number. These values are highly uncertain, but assuming a canonical value of $Q_\star/k_\star = 10^6$, we find that the decay time is $\approx 100$~Myr. If, however, $Q_\star/k_\star = 10^2$, as \cite{schlaufman2013} suggest may be the case for subgiant stars, then $t \approx 10,000$~yr. With such a short timescale, it would be an interesting coincidence to observe the planet in its current state. Interestingly, Kepler-91b \citep[e.g.][]{lillobox2014} has a tidal decay time scale that is of the same order of magnitude, but slightly shorter, because it orbits at a slightly higher semi-major axis but around a more evolved, larger star. Consequently, the existence of K2-39b and Kepler-91b appears to argue against the strong tidal dissipation suggested by \cite{schlaufman2013} to explain the under-abundance of short-period planets orbiting subgiant stars.

K2-39b may allow a direct test of the tidal dissipation strength in the future. Because $t = a/\dot{a} = P/\dot{P}$, we find that $\dot{P} = -4$~ms/yr for $Q_\star/k_\star = 10^6$, and $\dot{P} = -40$~s/yr for $Q_\star/k_\star = 10^2$. Very recently, $\dot{P} = (-2.56 \pm 0.40) \times 10^{-2}$ s~yr$^{-1}$ was measured for WASP-12b, based on ten years of transit observations, corresponding to a tidal quality factor of $2.5 \times 10^5$ for the (main-sequence) host star \citep{maciejewski2016}.
 
To aid future measurements of $\dot{P}$ for K2-39b, we report the times of the 15 individual transits observed by K2 in Table~\ref{tab:transittimes}. These times were measured by fitting the best transit model to individual transit observations, while the uncertainties were estimated through a bootstrap procedure, in which the residuals after the fit were resampled. The times and uncertainties reported in Table~\ref{tab:transittimes} are the mean and standard deviation of 4000 such fits to each transit. 

We also fitted the current transit times to place an upper limit on period decay. Modeling the time of each transit ($T_n$) as 
\begin{equation}
 T_n = T_0 + n P + \frac{1}{2}n^2 P\dot{P},
\end{equation}
we fit for $T_0$, $P$ and $\dot{P}$ using an MCMC algorithm \citep{foremanmackey2013}, with uniform priors on $T_0$ and $\dot{P}$ and a Gaussian prior on $P$ based on the simultaneous transit and RV fit reported in Table~\ref{tab:parameters}. Within 95\% confidence, we find that $\dot{P} > -0.000071$, corresponding to a period decay less than 37~min/yr. 

This provides a weak lower limit of $Q_\star/k_\star >\approx 1.8$. A longer baseline of observations could improve this constraint by orders of magnitude. Given the transit depth of $\approx~400$~ppm, observing future transits is difficult to do using ground-based observations. However, the TESS mission \citep[][]{ricker2014}, planned to observe in $\sim$2018-2019, or the CHEOPS mission \citep{broeg2013}, planned to observe in 2018-2020, should easily be able to observe the transits if they target this star. By this time, tidal strengths suggested by \cite{schlaufman2013} could lead to a period decay of several minutes, which should be well within reach of detectability. 

Finally, we note that as K2 continues to observe, it may discover other rare systems similar to K2-39, allowing us to further constrain stellar structure and planet formation and evolution.

\begin{table}[h]
\caption{Times of individual transits.\label{tab:transittimes}}
\begin{center}
 \begin{tabular}{c} %r@{$\pm$}l c}
       \tableline\tableline
      \noalign{\smallskip}
      Time [BJD]\\ %&  \multicolumn{2}{c}{RV [m s$^{-1}\mathrm{]}$}	& Instrument\\
      \noalign{\smallskip}
      \hline
2456980.8237	$\pm$ 	0.0076\\
2456985.4232	$\pm$	0.0081\\
2456990.010	$\pm$	0.011\\
2456994.6438	$\pm$	0.0086\\
2456999.2504	$\pm$	0.0093\\
2457003.899	$\pm$	0.011\\
2457008.4569	$\pm$	0.0074\\
2457013.0664	$\pm$	0.0083\\
2457017.6455	$\pm$	0.0093\\
2457022.2631	$\pm$	0.0081\\
2457026.8919	$\pm$	0.021 \\
2457031.475	$\pm$	0.010\\
2457036.084	$\pm$	0.017\\
2457040.715	$\pm$	0.020\\
2457045.292	$\pm$ 	0.013\\
      \noalign{\smallskip}
 \end{tabular}
 \end{center}
\end{table}

% header: system, time (bjd), rv (m/s), rv sig(m/s), fwhm (km/s), bis
% (km/s), spectrograph
\begin{table*}[htbp]
\caption{Radial velocity observations. \label{tab:rvdata}}
\begin{center}
 \begin{tabular}{ cccccc } %{c r@{$\pm$}l c}
       \tableline\tableline
      \noalign{\smallskip}
      Time [BJD] &  RV [m s$^{-1}$]	& $\sigma_{\mathrm{RV}}$ [m s$^{-1}$]	& FWHM [km s$^{-1}$] & BIS [km s$^{-1}$] & Instrument\\
      \noalign{\smallskip}
      \hline
  $2457255.71433$  &  $  24507.93$&$  2.66$&$  6.823$&$   0.053$ & HARPS \\
  $2457256.65992$  &  $  24505.27$&$  2.62$&$  6.813$&$   0.051$ & HARPS \\
  $2457257.78349$  &  $  24485.83$&$  1.31$&$  6.809$&$   0.047$ & HARPS \\
  $2457258.70438$  &  $  24462.17$&$  2.40$&$  6.813$&$   0.048$ & HARPS \\
  $2457260.71027$  &  $  24501.40$&$  2.28$&$  6.817$&$   0.049$ & HARPS \\
  $2457276.75557$  &  $  24473.82$&$  3.85$&$  6.814$&$   0.048$ & HARPS \\
  $2457278.74939$  &  $  24500.67$&$  1.88$&$  6.799$&$   0.054$ & HARPS \\
  $2457235.66962$  &  $  24557.22$&$  6.99$&$  11.538$&$ -0.021$ & FIES \\
  $2457239.50242$  &  $  24580.43$&$  5.65$&$ 11.549$&$   0.003$ & FIES \\
  $2457239.57252$  &  $  24582.36$&$  5.52$&$  11.560$&$ -0.013$ & FIES \\
  $2457239.65383$  &  $  24576.87$&$  6.13$&$  11.547$&$ -0.011$ & FIES \\
  $2457240.51866$  &  $  24581.82$&$  4.81$&$  11.558$&$ -0.003$ & FIES \\
  $2457240.62475$  &  $  24575.68$&$  5.84$&$  11.551$&$ -0.020$ & FIES \\
  $2457241.55160$  &  $  24590.82$&$  7.26$&$  11.573$&$ -0.002$ & FIES \\
  $2457241.64591$  &  $  24595.43$&$  5.24$&$ 11.546$&$   0.005$ & FIES \\
  $2457248.67838$  &  $  24565.10$&$ 18.44$&$ 11.552$&$   0.015$ & FIES \\
  $2457249.58312$  &  $  24575.34$&$  6.52$&$  11.548$&$ -0.006$ & FIES \\
  $2457261.53317$  &  $  24577.38$&$  6.00$&$  11.545$&$ -0.008$ & FIES \\
  $2457262.66662$  &  $  24554.68$&$  7.26$&$  11.578$&$ -0.009$ & FIES \\
  $2457342.42953$  &  $  24539.24$&$  6.00$&$ 11.546$&$   0.003$ & FIES \\
  $2457343.42036$  &  $  24556.29$&$  9.07$&$ 11.531$&$   0.018$ & FIES \\
  $2457344.41133$  &  $  24544.61$&$  7.31$&$  11.559$&$ -0.014$ & FIES \\
  $2457394.32683$  &  $  24583.37$&$  5.60$&$  11.556$&$ -0.020$ & FIES \\
  $2457395.31999$  &  $  24581.37$&$  5.30$&$ 11.546$&$   0.004$ & FIES \\
  $2457257.79909$  &  $      0.73$&$  1.64$&  -- &-- & PFS \\
  $2457258.77452$  &  $     -9.97$&$  1.57$&--&-- & PFS \\
  $2457261.80523$  &  $      0.00$&$  1.46$&--&-- & PFS \\
  $2457267.70240$  &  $    -20.64$&$  1.40$&--&-- & PFS \\
  $2457268.78307$  &  $     -5.45$&$  1.51$&--&-- & PFS \\
  $2457269.72528$  &  $     14.95$&$  1.54$&--&-- & PFS \\
 \end{tabular}
 \end{center}
\end{table*}

\acknowledgements

{\small We thank the referee, Alexander Santerne, for helpful comments and suggestions which significantly improved this manuscript. We thank Saul Rappaport for helpful comments during the early stages of this project. We acknowledge kind help by Masayuki Kuzuhara for the analysis of Subaru IRCS data.
N.N. acknowledges support by the NAOJ Fellowship, Inoue Science Research Award, and Grant-in-Aid for Scientific Research (A) (No. 25247026) from the Ministry of Education, Culture, Sports, Sci- ence and Technology (MEXT) of Japan.
I. R. acknowledges support from the Spanish Ministry of Economy and Competitiveness (MINECO) and the Fondo Europeo de Desarrollo Regional (FEDER) through grants ESP2013-48391-C4-1-R and ESP2014-57495-C2-2-R. 
A.V. is supported by the NSF Graduate Research Fellowship, Grant No. DGE 1144152.
This work was performed [in part] under contract with the California Institute of Technology (Caltech)/Jet Propulsion Laboratory (JPL) funded by NASA through the Sagan Fellowship Program executed by the NASA Exoplanet Science Institute.
This article is based on observations obtained with the Nordic Optical Telescope, operated on the island of La Palma jointly by Denmark, Finland, Iceland, Norway, and Sweden, in the Spanish Observatorio del Roque de los Muchachos of the Instituto de Astrofisica de Canarias. Further observations made with the 1.55-m Carlos S\'anchez Telescope operated on the island of Tenerife by the Instituto de Astrof\i sica de Canarias in the Spanish Observatorio del Teide. Observations with the HARPS spectrograph at ESO's La Silla observatory (095.C-0718(A)). Data gathered with the 6.5 meter Magellan Telescopes located at Las Campanas Observatory, Chile. Funding for the Stellar Astrophysics Centre is provided by The Danish National Research Foundation (Grant agreement no.: DNRF106). The research is supported by the ASTERISK project (ASTERoseismic Investigations with SONG and Kepler) funded by the European Research Council (Grant 
agreement no.: 267864). We acknowledge ASK for covering travels in relation to this publication. The research leading to these results has received funding from the European Union Seventh Framework Programme (FP7/2013-2016) under grant agreement No. 312430 (OPTICON). This research has made use of the Exoplanet Orbit Database and the Exoplanet Data Explorer at exoplanets.org.}

%\clearpage

%\clearpage
\bibliographystyle{bibstyle}
\bibliography{references_k2}

\end{document}